\documentclass[12pt]{article}
\usepackage[utf8]{inputenc}
\usepackage[export]{adjustbox}
\usepackage{authblk}
\usepackage{bbm}
\usepackage{pgfpages}
\usepackage{graphicx}
\usepackage{multicol}
\usepackage{csquotes}
\usepackage{multirow}
\usepackage{graphicx} 
\usepackage{booktabs} 
\usepackage{authblk}
\usepackage{caption}
\usepackage{subcaption}
\usepackage{stackengine}
\usepackage{amsmath}
\PassOptionsToPackage{hyphens}{url}\usepackage{hyperref}
\usepackage{comment}
\usepackage[capposition=top]{floatrow}
\usepackage{hyperref}
\usepackage{lscape}
\hypersetup{
    colorlinks=true,
    linkcolor=blue,
    filecolor=blue,      
    urlcolor=blue,
    citecolor=blue,
}
\usepackage[capposition=top]{floatrow}

\usepackage[spanish,english]{babel}

\usepackage{amssymb}
\usepackage{stackrel}
\usepackage{natbib} 

\usepackage[a4paper, margin=0.75in]{geometry}

\title{International Spillovers of ECB Interest Rates \\ \Large Monetary Policy \& Information Effects} 

\author[1]{Santiago Camara\footnote{Contact email: santiagocamara2022@u.northwestern.edu .}}

\affil[1]{Northwestern University}

\date{\today}

\begin{document}
    
\date{}
\maketitle
\begin{center}
\normalsize
This version: \today. For the latest version \href{https://scamara91.github.io/JMP/ECBSpilloversBias.pdf}{Click here!}
\end{center}

\begin{abstract}
This paper shows that disregarding the information effects around the European Central Bank monetary policy decision announcements biases its international spillovers. Using data from 23 economies, both Emerging and Advanced, I show that following an identification strategy that disentangles pure monetary policy shocks from information effects lead to international spillovers on industrial production, exchange rates and equity indexes which are between 2 to 3 times larger in magnitude than those arising from following the standard high frequency identification strategy. This bias is driven by pure monetary policy and information effects having intuitively opposite international spillovers. Results are present for a battery of robustness checks: for a sub-sample of ``close'' and ``further away'' countries, for both Emerging and Advanced economies, using local projection techniques and for alternative methods that control for ``information effects''. I argue that this biases may have led a previous literature to disregard or find little international spillovers of ECB rates.

\medskip

\medskip


\medskip

\end{abstract}

\newpage

\section{Introduction} \label{sec:introduction}

High-frequency changes in interest rates around the Federal Reserve’s (Fed) and the European Central Bank’s (ECB) policy announcements are a standard method of measuring monetary policy shocks (standard HFI strategy). However, there is a substantial literature which has found puzzling effects of these shocks in domestic variables, such as GDP, inflation, and interest rates. This evidence has been supportive of a ``information effects’’ or an ``information channel'' of monetary policy, whereby a Fed or ECB tightening may communicate that the economy is stronger than the private sector expected. While there is evidence that information effects may cause puzzling results when estimating the international spillovers of the Fed’s interest rates, there is little to no evidence on the role informational effects have on shaping the ECB’s international spillovers.
This paper shows that disregarding the information effects around ECB policy announcements biases the estimates of the international spillovers of ECB’s interest rates. Following an identification strategy that disentangles pure monetary policy shocks from information disclosure shocks lead to an impact in industrial production and nominal exchange rates which is between 2 to 3 times larger in magnitude than following the standard high frequency identification strategy. Furthermore, disregarding the information effect would lead to the conclusion that an ECB tightening causes an expansion of equity indexes abroad. This bias is driven by pure monetary policy and information disclosure shocks having intuitively opposite international spillovers.

I show this by estimating a panel SVAR model using an identification strategy scheme that allows me to separate two ECB shocks: a pure monetary policy shock (MP shock) and an information disclosure shock (ID shock). I disentangle these two shocks using the methodology introduced by \cite{jarocinski2020deconstructing} which imposes sign restrictions on the co-movement of the high-frequency surprises of interest rates and the stock market around ECB meetings. This co-movement is informative as standard theory unambiguously predicts that a monetary policy tightening shock should lead to lower stock market valuation. This is because a pure monetary policy tightening decreases the present value of future dividends by increasing the discount rate and by deteriorating present and future firm’s profits and dividends. Thus, MP shocks are identified as those innovations that produce a negative co-movement between these high-frequency financial variables. On the contrary, innovations generating a positive co-movement between the interest rates and the stock market correspond to ID shocks. This is a shock that occurs systematically at the time of the ECB policy announcements, but that is different from the standard monetary policy shock. By separately identifying these two structural shocks and introducing them into a panel SVAR, I find that MP shocks produce a recession, an exchange rate depreciation and a drop in equity indexes. On the contrary, ID shocks produce an economic expansion, an exchange rate appreciation and an expansion of equity indexes.

I show that results are robust across model and sample specifications. First, I find that the biases arising from disregarding the information effects of ECB interest rates are greater for countries further away from the Euro Area. This piece of evidence suggests that following the standard HFI strategy may lead to the wrong conclusion that ECB interest rates have little to no quantitative international spillovers. Second, I show results are present for both Emerging and Advanced economies. Third, I show that results also arise when estimating the impulse response functions using local projection regressions. Lastly, I show that results also arise when using alternative identification strategies that control for ``information effects''.

\noindent
\textbf{Related literature.} This paper relates to two main strands of literature. First, this paper contributes to an increasingly larger literature which has studied the role of information effects around the Fed’s and the ECB’s policy meetings such as \cite{romer2000federal}, \cite{nakamura2018high} and \cite{jarocinski2018macroeconomic}, which argue that these central banks have considerable better information about aggregate variables than commercial/professional forecasters, and  that policy actions provide information signals that lead to private sector agents revising their beliefs and expectations over aggregate dynamics. More recently, \cite{jarocinski2020deconstructing} introduces an identification strategy disentangles these monetary policy and the information components of monetary policy by studying high-frequency movements in interest rates and stock prices around policy announcements. The authors show that pure monetary policy and information disclosure shocks have intuitive and very different effects on the US economy. Additionally, the authors argue that ignoring the central bank information shocks biases the inference on monetary policy non-neutrality within the US economy. The contribution of this paper to this literature is two fold. First, providing evidence of significant information effects around ECB policy announcements which cause substantial spillovers the rest of the world. Second, this paper focuses on showing how disregarding the systematic disclosure of information around policy meetings consistently biases the international spillovers of ECB interest rates and may have lead previous research to conclude the lack of any significant international transmission.\footnote{For a discussion on the small or lack of international transmission of ECB interest rates see \cite{ca2020monetary}.}

Additionally, this paper contributes to a long strand of literature which has focused on quantifying the international spillovers of foreign interest rate shocks into open economies, such as \cite{eichenbaum1995some} and \cite{uribe2006country} for the Fed, and \cite{von1990german} and \cite{kobrak2008banking} for the ECB. More recently, \cite{jarocinski2022central} studies the impact of Fed and ECB MP and ID shocks in the US and Euro Area economies, finding that while Fed shocks have spillovers the Euro Area, ECB MP shocks do not have significant effects in the US economy. Additionally, \cite{miranda2022tale} shows that ``quantitative easing'' shocks by the ECB have significant global spillovers, though relatively smaller than those by the Fed. In this paper, I show that while ECB MP shocks may not have a significant impact on the US economy, both MP and ID shocks have quantitatively large impacts for both Emerging and Advanced economies. Closely related to this paper, \cite{camara2021spillovers} shows that following the standard HFI leads to puzzling international spillovers, such as an economic expansion, an exchange rate appreciation and equity index booms, which disappear once one controls for the information disclosed around ECB meetings.\footnote{Also related to this paper, \cite{camararamirez2022} shows that Fed and ECB shocks affect the Chilean economy through different transmission channels by exploiting firm level balance sheet data.} The contribution of the this paper to this literature are two-fold. To begin with, this paper shows that ECB interest rates have quantitatively meaningful international spillovers in open economies, highlighting the global role of the ECB and the Euro Area in financial and international trade markets. Unlike \cite{jarocinski2022central} and \cite{miranda2022tale}, this paper uses detailed data from a panel of 23 countries, both Emerging Markets and Advanced Economies, instead of relying on US, Euro Area or global variables.\footnote{It is noteworthy to mention that \cite{miranda2022tale} that estimates the impact of a ``Path Factor'' of ECB shocks to measure the informational impact of ECB's quantitative easing. In that paper, the authors follow the identification strategy proposed by \cite{swanson2021measuring} which exploits different interest rate maturities such that the ``Path Factor'' proxies for shocks that affect markets expectations of future medium-term interest rates which combines the effects of forward guidance with those of this ``signalling channe'' of the QE transmission mechanism. On the contrary, in this paper I use the identification strategy proposed by \cite{jarocinski2020deconstructing} which exploits the co-movement of both interest rates and the stock market to isolate a pure information effect.} Secondly, this paper shows that failing to take into account the information effects around ECB meetings leads to downward-biased estimates of the impact of ECB interest rates into the rest of the world. Thus, I argue that the biases introduced by information effects may explain why a previous research found little to no international spillovers of ECB interest rates.\footnote{For a discussion on the lack of international spillovers of ECB interest rates see \cite{ca2020monetary}.}

\section{Data, Methodology \& Identification} \label{sec:data_methodology_identification}

In this section, I describe the construction of my dataset, describe the panel SVAR methodology used and delineate the identification strategy used across the paper.

\noindent
\textbf{Data.} First, I describe the sample of Emerging Economies (EM) and Advanced Economies (AE) countries, the different datasets used across the paper to construct out sample of macroeconomic and financial variables, and the source of the high-frequency surprises and ECB shocks. The benchmark specification analyzes the international transmission of ECB interest rate spillovers on 23 countries, 16 Emerging Markets and 7 Advanced Economies at the monthly frequency for the period January 2004 to December 2016. The Emerging Market Economies are: Brazil, Bulgaria, Chile, Colombia, Hungary, India, Indonesia, Malaysia, Mexico, Peru, Philippines, Poland, Russia, South Africa, Turkey, Uruguay. The Advanced Economies are: Australia, Canada, Iceland, Japan, Singapore, South Korea and Sweden.\footnote{ A big constraint in the construction of a rich but balanced panel dataset is the lack of data availability for Emerging Market economies in the late 1990s and early 2000s. It is noteworthy to mention that during the late 1990s and early 2000s several EM economies experienced significant monetary and fiscal policy changes (for instance the implementation of inflation targetting regimes and fiscal policy rules). An example of these policy changes is that around a third of emerging and developing countries shifted from pro-cyclical to counter-cyclical fiscal policies between the late 1990s to the early 2000s.} 

The benchmark model specification includes four macroeconomic and financial variables: (i) nominal exchange rate with respect to the Euro, (ii) industrial production index, (iii) CPI index, (iv) equity index.\footnote{In order to construct a harmonized dataset of macroeconomic and financial variables for both AE and EM economies, I source all of the datasets from the IMF, which guarantees that the variables in the dataset are constructed following closely related methodologies. Variables ``industrial production'', ``consumer price index'', ``nominal exchange rate with respect to the Euro'', ``equity index'' are sourced from the IMF's International Financial Statistics dataset. For additional details on the construction of the dataset see Appendix \ref{sec:appendix_data_details}.} I source the high frequency surprises in interest rates and the two ECB shocks from the dataset constructed by \cite{jarocinski2022central}.$^{,}$\footnote{Both the time series and the the structural shocks and the replication codes to compute shocks can be directly downloaded from the authors' website. See \url{https://marekjarocinski.github.io/}.}

\noindent
\textbf{Methodology.} Next, I describe the panel SVAR model methodology estimated across the paper. The model specification is a pooled panel SVAR as presented by \cite{canova2013panel}, following \cite{camara2021spillovers}. This type of model considers the dynamics of several countries simultaneously, but assuming that the dynamic coefficients are homogeneous across units, and coefficients are time-invariant. In this framework, this implies that country $i$'s variables only depend on structural shocks and the lagged values of country $i$'s variables.

\noindent
\textbf{Identification strategy.} The identification strategy combines the two structural ECB shocks recovered by using the identification methodology developed by \cite{jarocinski2020deconstructing} and \cite{jarocinski2022central} with a standard Choleski ordering identification strategy. The identification strategy introduced by \cite{jarocinski2020deconstructing} exploit the high-frequency surprises of multiple financial instruments to recover two distinct ECB shocks: a pure monetary policy (MP) shock and an information disclosure (ID) shock. This deconstruction faces a typical set identification problem.\footnote{See \cite{jarocinski2020deconstructing} and \cite{jarocinski2022central} to see a detailed discussion of this topic.}. For my benchmark identification strategy I follow \cite{jarocinski2022central} and use the shock deconstruction that arises from using the median rotational angle that satisfies the sign restriction condition.\footnote{In Section \ref{subsec:robustness_checks_additional_results} I show that the benchmark results are robust to using different rotational angles.} I order the vector of identified structural shocks first with the vector of country $i$ specific macroeconomic and financial variables second. I introduce the high-frequency surprise in interest rates and the two ECB shocks one at a time and compute the resulting impulse response functions for each shock separately. In Section \ref{subsec:robustness_checks_additional_results} I show that an alternative method to control for information effects around ECB meetings yields similar results.

\section{Spillovers of Monetary Policy \& Information Effects} \label{sec:main_results}

This section presents the main results of this paper. In Section \ref{subsec:benchmark_results} I estimate and quantify the impact of the two ECB shocks: a pure monetary policy (MP) shock and an information disclosure (ID) shock; and following the standard HFI. I show that the two ECB shocks have completely opposite international spillovers. I argue that the presence of information disclosure shocks biases the results arising from the standard HFI strategy. In Section \ref{subsec:robustness_checks_additional_results} I show that my results are robust across different sub-samples, to estimating the impulse response functions using local projection regressions, and to alternative identification strategies that control for ``information effects'' around ECB meetings.

\subsection{Benchmark Results} \label{subsec:benchmark_results}

\begin{figure}[ht]
         \centering
         \caption{Impulse Response to One-Standard-Deviation Shock \\ \footnotesize Benchmark Specification}
         \includegraphics[scale=0.425]{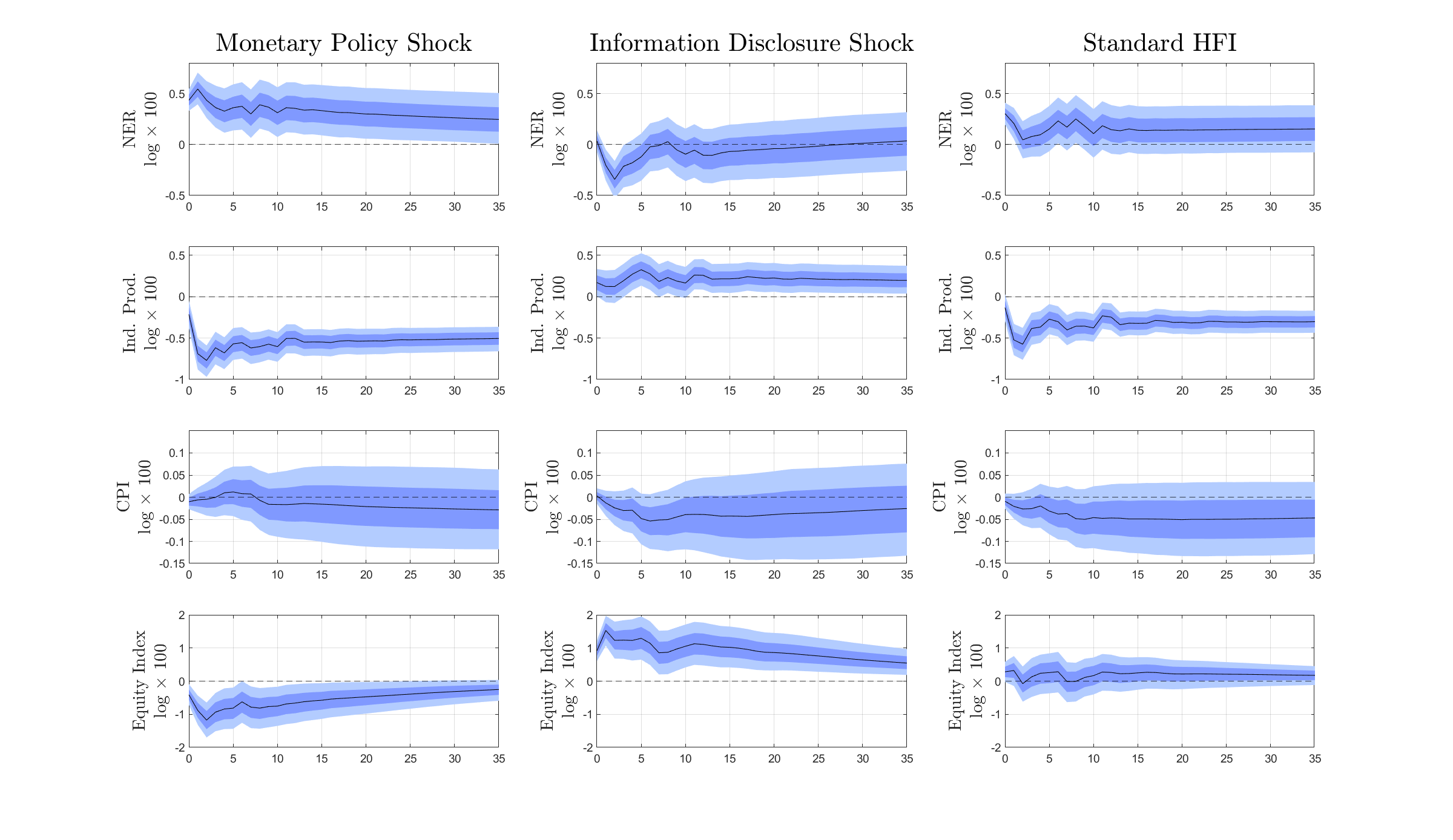}
         \label{fig:Benchmark}
         \floatfoot{\scriptsize \textbf{Note:} The black solid line represents the median impulse response function. The dark shaded area represents the 68 confidence intervals. The light shaded area represents 95 confidence intervals. The figure is comprised of 12 sub-figures ordered in four rows and three columns. Every row represents a different variable: (i) nominal exchange rate, (ii) industrial production index, (iii) consumer price index, (iv) equity index. The first column presents the results for the MP or ``Pure Monetary Policy'' shock, the middle column presents the results for the ID or ``Information Disclosure'' shock, and the last column presents the results for the interest rate composite high frequency surprise or ``Standard HFI''. In the text, when referring to Panel $(i,j)$, $i$ refers to the row and $j$ to the column of the figure.}
\end{figure}

I start by showing that deconstructing interest rate movements into the two distinct structural shocks matter for quantifying the international spillovers of ECB interest rates. The first and second columns of Figure \ref{fig:Benchmark} present the impulse response functions of the macroeconomic and financial variables to a MP and ID shock, respectively.
Comparing the figures across these two columns leads to a first important conclusion. This is that the identification scheme based on sign restriction of the policy rates and the stock market financial separately identifies two distinct economic shocks. If the co-movement between the high frequency surprises of the policy interest rate and the stock market was uninformative, the impulse response functions presented in the first and second columns of Figure \ref{fig:Benchmark} should exhibit the same results. Comparing the results presented in these figures, it is straightforward to conclude that this is not the case. For instance, the behavior of the industrial production and equity indexes are completely opposite across figures, with a MP shock leading to a persistent decline in both indexes and a ID shock leading to a persistent increase in them. Hence, correctly identifying the different EBC shocks is crucial to accurately quantify the spillovers of ECB monetary policy shocks.

I now turn to describing the impulse response functions in greater detail. The first column of Figure \ref{fig:Benchmark} shows the impulse responses to a MP shock. First, Panel (1.1) shows that a one-standard-deviation MP shock leads to a 50 basis point depreciation on impact of the euro nominal exchange rate, remaining significantly above pre-shock levels for at least 35 months after the shock. Panel (2,1) shows that a MP shock leads to hump shaped reduction in industrial production. After a 25 basis point decrease on impact, industrial production continues decreasing and reaches a trough 80 basis points below pre-shock level 3 months after the initial shock. The decrease in industrial production is persistent, remaining 50 basis points below pre-shock levels 35 months after the initial shock. Panel (3,1) shows that a MP shock does not lead to a significant response of the consumer price index. This lack of exchange rate pass-through may be explained by the increasing evidence dollar currency pricing, i.e., international trade invoiced in US dollars, which could reduce the influence of euro exchange rates on domestic pricing. Lastly, Panel (4.1) shows that a ECB MP shock leads to a hump shaped drop in the equity index, close to 50 basis points on impact and reaching a trough 100 basis points below pre-shock levels. 

The international spillovers of an ECB ID shock are completely opposite to those of a MP shock. The second column of Figure \ref{fig:Benchmark} presents the impulse response functions of a one-standard-deviation ID shock. Panel (1.2) shows that the nominal euro exchange rate shows an appreciation in response to an ID shock. However, this appreciation is short-lived, only significantly different from zero during the first 4 months after the initial shock. Panel (2.2) shows that industrial production exhibits a mild yet persistent increase in response to a ID shock. The increase becomes significantly different from zero 4 months after the initial shock and remains 25 basis points above pre-level shocks for 35 months. Panel (3.2) shows that a ID shock leads to a reduction in the consumer price index, however this is only significantly different from zero when compared to a 68\% confidence interval, but not compared to a 95\% confidence interval. Lastly, Panel (4.2) shows that a ID shocks leads to a on impact jump in the equity index of 100 basis points. Additionally, the index continues increasing and reaching a peak of 150 basis points above the pre-shock level 2 months after the initial shock. In summary, the first and second columns of Figure \ref{fig:Benchmark} show that the two EBC shocks lead to almost completely opposite impulse response functions for the full sample of countries. 

Next, I argue that following the ``Standard HFI'' strategy, i.e., using only the high-frequency surprise of the policy interest rate composite, leads to biased estimates of the ECB's international spillovers. The third column of Figure \ref{fig:Benchmark} exhibits the impulse response functions under this identification strategy. Across the different variables, the impulse responses are an average of the responses presented for the MP and ID shocks in the first and middle columns. Comparing the results for the nominal euro exchange rate, Panel (1.3) shows that following the standard HFI approach would lead to the conclusion that an ECB interest rate hike has no significant impact on nominal exchange rates. In terms of impact on industrial production, following the standard HFI approach, Panel (2.3), leads to, on average, a 40\% smaller impact than when purging the information effects, see Panel (2.1). For instance, following the MP shock (Standard HFI shock) the drop in industrial production on impact is 21 basis points (13 basis points), reaches a trough of 68 basis points (38 basis points, and converges to a value 50 basis points (30 basis points) below pre shock levels. Finally, the impact on the equity index, on the fourth row, provides the sharpest qualitative difference. Following the standard HFI approach, Panel (4.3), yield a mild expansion in the equity index. However, once controlling for the information effects around policy announcements shows that an ECB interest rate hike caused by a monetary policy shock, Panel (4.1) leads to a significant and persistent drop in equity indexes.  

\noindent 
\textbf{Discussion.} Overall, the results presented above show that disregarding the systematic disclosure of information around policy announcements biases the international spillovers of ECB interest rates. This is because an interest rate hike caused by the disclosure of positive information about the state of the Euro Area economy around ECB policy announcements, i.e. a ID shock, leads to opposite international spillovers than those generated by a pure monetary policy shock, i.e. MP shock. Thus, following the standard HFI approach could lead to the erroneous conclusion that ECB interest rates do not have significant international spillovers.

The fact that information effects around ECB policy announcements biases international spillovers of its interest rates is line with what \cite{camara2021spillovers} found for Fed interest rates. Yet, the biases introduced by information effects are quantitatively smaller for the case of ECB than for Fed interest rates.\footnote{For instance, \cite{ilzetzki2021puzzling} and \cite{camara2021spillovers} show that following the standard HFI approach lead to an expansion of industrial production and looser financial conditions in the rest of the world.} This is expected as, for instance, \cite{nakamura2018high} showed for the US that following the standard HFI approach leads to an increase in real interest rates, expected inflation, and expected output growth simultaneously, which the authors attribute to changes in private sector beliefs about both monetary policy but also about other economic fundamental variables. Additionally, this relatively smaller role of information effects around ECB policy announcements is in line with evidence presented by \cite{jarocinski2022central}, which presented evidence that information effects bias the degree of monetary non-neutrality within the Euro Area, but do not change the sign of impact, as presented for the US by \cite{nakamura2018high}.  

\subsection{Additional Results \& Robustness Checks} \label{subsec:robustness_checks_additional_results}

In this section, I present additional results and robustness checks that complement the benchmark results presented above. For convenience of presentation, all figures are presented in Appendix \ref{sec:appendix_additional_graphs}.

\noindent
\textbf{Results across different sub-samples.} Given the large number of countries and their different underlying characteristics, a natural robustness check exercise is to test whether the results presented above hold across different country sub-samples. I carry out two sub-sample exercises: (i) partitioning the sample across the geographical distance into a sub-sample of ``close'' and a sub-sample of ``further-away'' countries to the Euro Area; and (ii) Emerging vs Advanced economies. The main results presented in Section \ref{subsec:benchmark_results} are robust across all two sub-sample exercises. Additionally, I discuss any minor quantitative differences across sub-samples.

Figure \ref{fig:Benchmark_Regions} in Appendix \ref{sec:appendix_additional_graphs} presents impulse response functions estimated for a sub-sample of economies geographically ``close'' to the Euro Area (Figure \ref{fig:Benchmark_CLOSER}), and a sub-sample of economies ``further-away'' from the Euro Area (Figure \ref{fig:Benchmark_FAR}).\footnote{The sub-sample of economies ``close'' to the Euro Area is: Bulgaria, Hungary, Iceland, Poland, Russia, Sweden and Turkey. The sub-sample of economies ``further-away'' from the Euro Area is: Australia, Brazil, Canada, Chile, Colombia, India, Indonesia, Japan, Korea, Malaysia, Mexico, Peru, Philippines, Singapore, South Africa, Uruguay.} The first conclusion from this exercise is that our benchmark results are present in both sub-samples, i.e., pure monetary policy and information disclosure shocks have qualitatively different international spillovers and following the standard HFI approach would predict significantly lower impact in the exchange rate, industrial production and the equity index. Nevertheless, the impact in industrial production of ECB shocks is quantitatively larger for the sub-sample of countries relatively ``closer'' to the Euro Area than for the sub-sample of countries ``further-away''. This is intuitive as geographically closer economies tend to trade relatively more with the Euro Area, use the euro relatively more for invoicing and as a financial/savings instrument. Still, the potential biases from disregarding the information effects around ECB meetings may be particularly larger for the sub-sample of countries ``further-away'' from the Euro Area. To see this, note that following the standard HFI approach leads to a persistent increase in the equity markets in this sub-sample, which is completely explained by the information disclosure shock.

Figure \ref{fig:Benchmark_EMsVsAdv} in Appendix \ref{sec:appendix_additional_graphs} presents the separately estimated impulse response functions for Emerging Market (see Figure \ref{fig:Benchmark_EM}) and Advanced Economies (see Figure \ref{fig:Benchmark_Adv}). Once again, the main results are robust to this sub-sample exercise, with only quantitative differences across sub-samples. First, the biases or loses of disregarding information effects around ECB meetings is significantly larger for Emerging Markets. To see this, note that following the standard HFI predicts almost no impact at all in Emerging Markets' industrial production, see Panel (2,3) in Figure \ref{fig:Benchmark_EM}. However, a MP shock leads to a persistent drop in industrial production that is significantly below pre-shock levels 36 months after the initial shock, see Panel (2,1) in Figure \ref{fig:Benchmark_EM}. Note that for Advanced Economies, both the standard HFI and the MP shock predict a persistent drop in industrial production, with the MP shock impulse response function being two and three times greater in magnitude than the one computed following the standard HFI approach. In consequence, once again, disregarding the systematic disclosure of information around ECB meetings may be the cause behind

\noindent
\textbf{Results using local projection regressions.} Next, I show that the main results that arise when estimating impulse response functions from a panel SVAR results also arise when using local projection regressions. In particular, I estimate the following regression given by Equation \ref{eq:LP_fixed_effects_date}.
\begin{align} 
    y_{i,t+h} = \beta^{Shock}_{h} i^{\text{Shock}}_t + \sum^{J_y}_{j=1} \delta^{j}_i y_{i,t-j} + \sum^{J_x}_{j=1} \alpha^{j}_i x_{i,t-j} + \sum^{J_i}_{j=1} \left( \phi^{j}_i i^{\text{Shock}}_{t-j} \right) +\Gamma_i+\Gamma_{t} + \epsilon_{i,t}  \label{eq:LP_fixed_effects_date}
\end{align}
where $y_{i,t+h}$ is the outcome variable of interest of country $i$ in period $t+h$, $i^{\text{Shock}}_t$ is either the high-frequency surprise in the interest rate, or one of the two ECB shocks; $y_{i,t-j}$, $x_{i,t-j}$ and $i^{Shock}_{t-j}$ represent lagged valued of the variable of interest, other macro-finance control variables and the specific shock, respectively; $\Gamma_{i}$ represents a country fixed effect and $\Gamma_t$ represents a linear trend. Parameter $\beta^{Shock}_h$ measures the impact that shock $i^{\text{Shock}}_t$ has on variable $y_{i,t+h}$ in period $t+h$. Standard errors are clustered at the ``time'' dimension given that the ECB shocks hit all countries in our sample at the same time. 

Figure \ref{fig:Benchmark_LP} in Appendix \ref{sec:appendix_additional_graphs} presents the estimated coefficients $\beta_h$ for the full sample of countries for the two ECB shocks and when following the Standard HFI approach. The local projection estimated impulse response functions are in line with the benchmark results. On the one hand, the ECB pure monetary policy leads to a depreciation of the domestic currency with respect to the Euro which lasts slightly more than a year; and a persistent drop in the industrial production and equity indexes. On the other hand, an ECB information disclosure shock leads to a short-lived exchange rate appreciation, a mild expansion in industrial production and a significant increase in equity indexes. Additionally, when following the Standard HFI approach, the estimated coefficients for the nominal exchange rate and the industrial production index are, on average, 35\% and 48\% smaller in magnitude than those estimated when using the ECB pure monetary policy shock. Moreover, as it was the case in Section \ref{subsec:benchmark_results}, following the Standard HFI approach predicts an ECB interest hike predicts either a positive or no significant impact in equity indexes, while a pure monetary policy shock predicts a significant decline. These results are robust to removing the country fixed effects (see Figure \ref{fig:Benchmark_LP_NOFE}) and the time trend (see Figure \ref{fig:Benchmark_LP_NOFE_NoTrend}). Additionally, Figures \ref{fig:Benchmark_LP_Closer} to \ref{fig:Benchmark_LP_Adv} show that all the sub-sample exercises are also robust to estimating impulse response functions using local projection techniques.

\noindent
\textbf{Results using alternative recovered ECB shocks.} Sign restriction identification strategies face a non-uniqueness problem, as mentioned in Section \ref{sec:data_methodology_identification}. To provide robustness to our benchmark exercise, which uses the median rotational angle which satisfies the sign-restrictions, I show all of our results are robust to using three alternative rotational angles: the 68th percentile angles (following \cite{jarocinski2022central} poor man strategy), and the 10th and 90th percentile angles which test upper and lower bounds. Figures \ref{fig:Benchmark_j} to \ref{fig:Benchmark_90} that the main benchmark results in Section \ref{subsec:benchmark_results}, are robust to using these alternative approaches to the non-uniqueness problem in sign identification strategies.

\noindent
\textbf{Results using alternative controls for information effects.} The literature on the identification of monetary policy shocks have proposed several theories on the ``information effects'' and different methods to control for them.\footnote{There is still no consensus on the literature about which is the correct or appropriate method to control for these ``information effects''. For instance, \cite{bauer2022reassessment} argues that what \cite{jarocinski2020deconstructing} label as ``information disclosure'' shocks can be explained by the ``Fed responding to News Channel''. However, \cite{jarocinski2022central} responds to this view and presents additional evidence favoring the results presented by \cite{jarocinski2020deconstructing}.} While this paper does not aim to contribute to this discussion, I show that all the benchmark results presented in Section \ref{subsec:benchmark_results} are still present when using an alternative identification strategy to that proposed by \cite{jarocinski2020deconstructing}. To do so, I follow \cite{miranda2022tale}'s approach on ``poor man's sign restriction'' using the identification strategy constructed by \cite{swanson2021measuring}.\footnote{It is noteworthy to point out that \cite{miranda2022tale}'s main focus point are ECB's quantitative easing policy shocks and measuring their impact on measures of global risk aversion.} This approach implies the construction of shocks to the ECB's target policy rate for all policy meetings (which I label in Figures as the \cite{swanson2021measuring} method), for the set policy meetings in which there is a negative co-movement of the policy rate and the stock market (which I label as ``cleansed monetary policy'' shock), for the set of policy meetings in which the co-movement is positive (which I label as ``information content'' shock).

Figure \ref{fig:Alternative_Identification} in Appendix \ref{sec:appendix_additional_graphs} presents the results of this robustness check. Comparing the resulting impulse response functions lead to the same three conclusions presented in the previous section: (i) the drop in industrial production and equity indexes is, on average, 15\% and 40\% greater once ``information content'' is cleansed; (ii) disregarding ``information contents'' would predict no depreciation of the nominal exchange rate in response to an ECB hike; (iii) ``information content'' shocks lead to intuitive opposite responses than ``cleansed monetary policy'' shocks, such as an exchange rate appreciation, and a mild expansion of industrial production and equity indexes. Figure \ref{fig:Alternative_LP} show that these results are robust to estimating the impulse response functions using local projection regressions. Even more, estimating a ``rate race'' impulse response functions between the ``cleansed'' and ``uncleansed'' shocks, by introducing them jointly into Equation \ref{eq:LP_fixed_effects_date}, only lead to intuitive and statistically significant dynamics for the ``cleansed'' one (see Figure \ref{fig:Rate_Race_LP} in Appendix \ref{sec:appendix_additional_graphs}). In summary, I interpret these results as further and supporting evidence that ``information effects'' around policy meetings bias the international spillovers of ECB interest rates.

\section{Conclusion} \label{sec:conclusion}

This paper studies the international spillovers of ECB's interest rates, highlighting potential biases that arise from disregarding the systematic disclosure of information around policy meetings. To do so, I use an identification strategy that uses multiple asset high-frequency surprises around ECB meetings to recover two distinct interest rate shocks, a pure monetary policy shock and information disclosure shock. Introducing these shocks into a panel SVAR model with data from 23 countries, these two interest rate shocks have qualitatively and intuitively different international spillovers. A pure monetary policy shock leads to an exchange rate depreciation with respect to the Euro, and a persistent drop in industrial production and equity indexes. Oppositely, an information disclosure shock leads to an exchange rate appreciation and expansions of industrial production. Lastly, I show that disregarding the information effects and following the standard high-frequency identification strategy leads to international spillovers that are between 35\% and 65\% smaller in magnitude. This finding may explain why previous research found little to no international spillovers of ECB interest rates.

Results are robust across a battery of tests, such as across several sub-samples, both for SVAR and local projection regressions, and alternative identification strategies that cleanse for ``information effects'' around policy meetings. Going forward, it is crucial to further study whether Fed and ECB interest rates' international spillovers are transmitted differently or through similar channels. 

\newpage
\bibliography{main.bib}

\newpage
\appendix

\section{Data Details} \label{sec:appendix_data_details}

In this section of the appendix I provide additional details on the construction of the sample used across the paper. The source of the macroeconomic and financial data used for the construction of the variables in the benchmark variable specification is the IMF's ``International Financial Statistics''.\footnote{To access the IMF's IFS datasets go to \url{https://data.imf.org/?sk=4c514d48-b6ba-49ed-8ab9-52b0c1a0179b}.}

First, the benchmark specification is comprised of five variables:
\begin{enumerate}
    \item Nominal Exchange Rate
    
    \item Industrial Production index
    
    \item Consumer Price Index

    \item Equity Index
\end{enumerate}
Next, I present additional details for the construction of each of the variables
\begin{itemize}
    \item \underline{Nominal Exchange Rate:} The variable's full name at the IMF IFS data set is ``Exchange Rates, National Currency Per U.S. Dollar, Period Average, Rate''. 
    
    \item \underline{Industrial Production Index:} In order to construct countries' ``Industrial Production Index'' I rely on three variables of the IMF IFS' dataset:
    
    \begin{itemize}
        \item Economic Activity, Industrial Production, Index
        
        \item Economic Activity, Industrial Production, Seasonally Adjusted, Index
        
        \item Economic Activity, Industrial Production, Manufacturing, Index
    \end{itemize}
    Ideally, I would  construct the variable ``Industrial Production Index'' by choosing only one of the variables mentioned above. However, this is impossible as countries do not report to the IMF all three of these variables for our time sample, rey 2004 to December 2016. For instance, Peru provides neither the ``Economic Activity, Industrial Production, Index'' nor the ``Economic Activity, Industrial Production, Seasonally Adjusted, Index'', but does provide the ``Economic Activity, Industrial Production, Manufacturing, Index''. Visiting Peru's Central Bank statistics website, there is no ``Industrial Production Index'', but there is an ``Industrial Production, Manufacturing Index'', which coincides with the variable reported as ``Economic Activity, Industrial Production, Manufacturing, Index'' to the IMF.
    
    In order to deal with this, I establish the following priority between the three IMF IFS variables: (i) ``Economic Activity, Industrial Production, Seasonally Adjusted, Index'' (ii) ``Economic Activity, Industrial Production, Index'', (iii) ``Economic Activity, Industrial Production, Manufacturing, Index''.
    
\item \underline{Consumer Price Index:} Data for all countries except Australia is constructed using the variable ``Prices, Consumer Price Index, All items, Index'' from IMF IFS data set. Australia does not report a monthly CPI series to the IMF-IFS data set. Furthermore, the Australian Bureau of Statistics provides only quarterly data on their consumer price index.\footnote{See \url{https://www.abs.gov.au/statistics/economy/price-indexes-and-inflation/consumer-price-index-australia/jun-2022}.} Thus, for the case of Australia I proxy the monthly consumer price index by using the ``Prices, Producer Price Index, All Commodities, Index''. Once again, given that the paper's main results are robust to the different exercises that partition the sample, I believe that using this proxy variable does not guide any of the of the results presented in the paper. 

\item \underline{Equity Index:} In order to construct countries' ``Equity Index'' I rely on two variables of the IMF IFS' dataset:
    
    \begin{itemize}
        \item Monetary and Financial Accounts, Financial Market Prices, Equities, Index
        
        \item Monetary and Financial Accounts, Financial Market Prices, Equities, End of Period, Index
    \end{itemize}
    
    I establish a priority: (i) ``Monetary and Financial Accounts, Financial Market Prices, Equities, Index'' (ii) ``Monetary and Financial Accounts, Financial Market Prices, Equities, End of Period, Index''. Again, the data coverage is not complete for all countries for the full sample period of January 2004 to December 2016
\end{itemize}

\newpage
\section{Model Details} \label{sec:appendix_model_details}

In this section of the appendix I provide additional details on the panel SVAR model and the identification strategy described in Section \ref{sec:data_methodology_identification}. Section \ref{subsec:appendix_model_details_panel_svar_model} presents details on the panel SVAR model while Section \ref{subsec:appendix_model_details_identification} presents additional technical details on the recovery of structural shocks. 

\subsection{Panel SVAR Model} \label{subsec:appendix_model_details_panel_svar_model}

In this section of the appendix I provide additional details on the estimation of the Structural VAR model presented in Section \ref{sec:data_methodology_identification}. In its most general form, a panel SVAR model comprises of N countries or units, $n$ endogenous variables, $p$ lagged values and $T$ time periods. The pooled panel SVAR model can be written as
\begin{align} \label{eq:pooled_estimator}
\begin{pmatrix}
y_{1,t} \\
y_{2,t} \\
\vdots  \\
y_{N,t}
\end{pmatrix}
&=C+
\begin{pmatrix}
A^1 \quad 0 \quad \cdots \quad 0 \\
0 \quad  A^1 \quad \cdots \quad 0 \\
\vdots \quad \vdots \quad \ddots \quad \vdots \\
0 \quad 0 \quad \cdots \quad A^1 
\end{pmatrix}
\begin{pmatrix}
y_{1,t-1} \\
y_{2,t-1} \\
\vdots  \\
y_{N,t-1}
\end{pmatrix}
+ \cdots \nonumber \\
\\
&+ \nonumber
\begin{pmatrix}
A^p \quad 0 \quad \cdots \quad 0 \\
0 \quad  A^p \quad \cdots \quad 0 \\
\vdots \quad \vdots \quad \ddots \quad \vdots \\
0 \quad 0 \quad \cdots \quad A^p 
\end{pmatrix}
\begin{pmatrix}
y_{1,t-p} \\
y_{2,t-p} \\
\vdots  \\
y_{N,t-p}
\end{pmatrix}
+
\begin{pmatrix}
\epsilon_{1,t} \\
\epsilon_{2,t} \\
\vdots \\
\epsilon_{N,t}
\end{pmatrix}
\end{align}
where $y_{i,t}$ denotes an $n \times 1$ vector of $n$ endogenous variables of country $i$ at time $t$ and $A^{j}$ is an $n \times n$ matrix of coefficients providing the response of country $i$ to the $j^{th}$ lag at period $t$. Note that by assuming that $A^j_1 = A^j_n = A^j$ for $j=1,\ldots,n$ implies the assumption that the estimated coefficients are common across countries. $C$ is a $Nn\times1$ vector of constant terms which are also assumed to be common across countries. Lastly, $\epsilon_{i,t}$ is an $n \times 1$ vector of residuals for the variables of country $i$, such that
\begin{align*}
    \epsilon_{i,t} \sim \mathcal{N}\left(0,\Sigma_{ii,t}\right)
\end{align*}
with 
\begin{align*}
    \epsilon_{ii,t} &= \mathbb{E} \left(\epsilon_{i,t} \epsilon_{i,t}' \right) = \Sigma_c \quad \forall i \\
    \epsilon_{ij,t} &= \mathbb{E} \left(\epsilon_{i,t} \epsilon_{j,t}' \right) = 0 \quad \text{for } i \neq j
\end{align*}
The last two equations imply that, as for the model's auto-regressive coefficients, the innovation's variance is equal across countries. 

The model described by Equation \ref{eq:pooled_estimator} is estimated using Bayesian methods. In order to carry out the estimation of this model I first re-write the model. In particular, the model can be reformulated in compact form as

\begin{align} \label{eq:model_compact_form}
\underbrace{\begin{pmatrix}
y_{1,t}' \\
y_{2,t}' \\
\vdots  \\
y_{N,t}'
\end{pmatrix}}_{Y_t, \quad N \times n}
&=
\underbrace{\begin{pmatrix}
y_{1,t-1}' \ldots y_{1,t-p}' \\
y_{2,t-1}' \ldots y_{2,t-p}' \\
\vdots  \ddots \vdots \\
y_{N,t-1}' \ldots y_{N,t-p}'
\end{pmatrix}}_{\mathcal{B}, \quad N \times np}
\underbrace{\begin{pmatrix}
\left(A^{1}\right)' \\
\left(A^{2}\right)' \\
\vdots \\
\left(A^{N}\right)'
\end{pmatrix}}_{X_t, \quad np \times n}
+
\underbrace{\begin{pmatrix}
\epsilon_{1,t}' \\
\epsilon_{2,t}' \\
\vdots \\
\epsilon_{N,t}'
\end{pmatrix}}_{\mathcal{E}_t, \quad N \times n}
\end{align}
or
\begin{align}
    Y_t = X_t \mathcal{B} + \mathcal{E}_t
\end{align}
Even more, the model can be written in vectorised form by stacking over the $T$ time periods 
\begin{align}
    \underbrace{vec\left(Y\right)}_{NnT \times 1} = \underbrace{\left(I_n \otimes X \right)}_{NnT \times n np} \quad \underbrace{vec\left(\mathcal{B}\right)}_{n np \times 1} \quad + \quad \underbrace{vec\left(\mathcal{E}\right)}_{NnT \times 1}
\end{align}
or
\begin{align}
    y = \Bar{X} \beta + \epsilon
\end{align}
where $\epsilon \sim \mathcal{N}\left(0, \Bar{\Sigma}\right)$, with $\Bar{\Sigma} = \Sigma_c \otimes I_{NT}$.

The model described above is just a conventional VAR model. Thus, the traditional Normal-Wishart identification strategy is carried out to estimate it. The likelihood function is given by
\begin{align}
    f\left(y | \Bar{X} \right) \propto |\Bar{\Sigma}|^{-\frac{1}{2}} \exp \left(-\frac{1}{2} \left(y - \Bar{X}\beta\right)' \Bar{\Sigma}^{-1} \left(y - \Bar{X}\beta\right) \right)
\end{align}
As for the Normal-Wishart, the prior of $\beta$ is assumed to be multivariate normal and the prior for $\Sigma_c$ is inverse Wishart. For further details, see \cite{dieppe2016bear}. All of the panel SVAR model computations are carried out using the BEAR Toolbox version 5.1.

\subsection{Identification Strategy \& Shock Recovery} \label{subsec:appendix_model_details_identification}

In this section of the appendix, I present additional details on the sign-restriction identification strategy presented in Section \ref{sec:data_methodology_identification} which recovers the two ECB structural shocks. As stated in Section \ref{sec:data_methodology_identification}, this identification strategy follows \cite{jarocinski2020deconstructing} and \cite{jarocinski2022central}.

The identification strategy introduced by \cite{jarocinski2022central} exploit the high-frequency surprises of multiple financial instruments to recover two distinct ECB shocks: a pure monetary policy (MP) shock and information disclosure (ID) shock. In particular, the authors impose sign restrictions conditions on the co-movement of the high-frequency surprises of interest rates and the Euro Stoxx 50 around ECB meetings. This co-movement is informative as standard theory unambiguously predicts that a monetary policy tightening shock should lead to lower stock market valuation. This is because a monetary policy tightening decreases the present value of future dividends by increasing the discount rate and by deteriorating present and future firm's profits and dividends. Thus, MP shocks are identified as those innovations that produce a negative co-movement between these high-frequency financial variables. On the contrary, innovations generating a positive co-movement between interest rates and the Euro Stoxx 50 correspond to ID shocks. 

The high frequency surprise in the policy interest rate, $i^{Total}$, can be decomposed as
\begin{align}
    i^{\text{Total}} = i^{\text{MP}} + i^{\text{ID}}
\end{align}
where $i^{\text{MP}}$ is negatively correlated with the high frequency surprise of the $Euro Stoxx 50$ ``$s$'', and $i^{\text{ID}}$ is positively correlated with the ``$s$''. As shown by \cite{jarocinski2022central}, the sign restriction recovery of the structural shocks must satisfy the following decomposition
\begin{align}
    M = UC 
\end{align}
where $U'U$ is a diagonal matrix, $C$ takes the form of
\begin{align} \label{eq:matrix_rotation_appendix}
C =
\begin{pmatrix}
    1 & c^{MP}<0 \\
    1 & c^{ID}>0 \\
\end{pmatrix}
\end{align}
where $M=(i^{Total},s)$ is a $T \times 2$ matrix with $i^{Total}$ in the first column, $s$ in the second; $U = (i^{\text{MP}},i^{\text{ID}})$ is a $T \times 2$ matrix with $i^{\text{MP}}$ in the first column and $i^{\text{ID}}$ in the second column; and $T$ denoting the time length of the sample. By construction, $i^{\text{MP}}$ and $i^{\text{ID}}$ are mutually orthogonal. Matrix $C$ captures how $i^{\text{MP}}$ and $i^{\text{ID}}$ translates into financial market surprises.

The decomposition in \ref{eq:matrix_rotation_appendix} is not unique. In terms of \cite{jarocinski2022central} there is a range of rotations of matrices $U$ and $C$ that satisfy the sign restrictions $c^{MP}<0$ and $c^{ID}>0$.

The matrices $U$ and $C$ are computed as
\begin{align}
    U &= QPD \\
    C &= D^{-1} P' R
\end{align}
where the matrices $Q,P,D,R$ are obtained in three steps. 

\begin{itemize}
    \item[1.] Decompose matrix $M = UC$ into two orthogonal components using a QR decomposition such that
\end{itemize}
\begin{align}
    M &= QR \\
    Q'Q &=\begin{pmatrix}
        1 & 0 \\
        0 & 1 \\        
    \end{pmatrix} \\
    R&=\begin{pmatrix}
        r_{1,1}>0 & r_{1,2} \\
        0 & r_{2,2}>0 \\        
    \end{pmatrix}
\end{align}

\begin{itemize}
    \item [2.] Rotate these orthogonal components using the rotation matrix
    \begin{align}
    P &= \begin{pmatrix}
       \cos \left(\alpha\right) & \sin \left(\alpha\right) \\
       - \sin \left(\alpha\right) & \cos \left(\alpha\right)
    \end{pmatrix}
    \end{align}
    To satisfy the sign restrictions use any angle $\alpha$ in the following range
    \begin{align*}
        & \alpha \in \left( \left(1-w\right) \times \arctan \frac{r_{1,2}}{r_{2,2}} , \frac{w \times \pi}{2} \right) \qquad \text{if } r_{1,2} >0 \\
        & \alpha \in \left( 0 , w \times \arctan \frac{-r_{2,2}}{r_{1,2}} \right) \qquad \text{if } r_{1,2} \leq 0 \\
    \end{align*}
    where $w$ is weight, between 0 and 1, scaling the rotation angle. Setting $w = 0.5$ implies the median rotation angle, assumption used under the benchmark specification
\end{itemize}
\begin{itemize}
    \item[3.] Re-scale the resulting orthogonal components with a diagonal matrix $D$ to ensure that they add up to the interest rate surprises $i^{\text{Total}}$. It is straightforward to show that
    \begin{align}
        D = \begin{pmatrix}
            r_{1,1} \cos \left(\alpha\right) & 0 \\
            0 & r_{1,1} \sin \left(\alpha\right)
        \end{pmatrix}
    \end{align}
\end{itemize}

The first robustness check in Section \ref{subsec:robustness_checks_additional_results} which proposes an alternative approach to deal with the non-uniqueness problem of sign-restriction identification follows the approach of \cite{jarocinski2022central}. In \cite{jarocinski2022central}, the angle of rotation $\alpha$ of matrix $P$ is pinned down following these steps
\begin{itemize}
    \item[1.] Construct ``poor man's sign restrictions'' shocks such that
    
    \begin{align*}
        i^{\text{Total}} &= i^{MP} \quad \& \quad i^{\text{ID}} = 0 \qquad \text{if } i^{Total} \times s \leq 0 \\
        i^{\text{Total}} &= i^{ID} \quad \& \quad i^{\text{MP}} = 0 \qquad \text{if } i^{Total} \times s > 0 \\
    \end{align*}
\end{itemize}
In the data, \cite{jarocinski2022central} finds that the poor man's monetary policy shocks account for 68\% of the variance of the ECB's total interest rate surprises, i.e,
\begin{align*}
    \frac{var \left(i^{\text{MP}}\right)}{var \left(i^{\text{Total}}\right)} = 0.68
\end{align*}
To pin down the decomposition,  \cite{jarocinski2022central} impose that, as in the ``poor man's sign restrictions'' case, $var \left(i^{\text{MP}}\right)/ var \left(i^{\text{MP}}\right) = 0.68$.
In Section \ref{subsec:robustness_checks_additional_results} I carry out several robustness checks that considering different rotational angles. 

\section{Additional Results Figures} \label{sec:appendix_additional_graphs}

In this Appendix of the paper I present the figures referred to in Section \ref{subsec:robustness_checks_additional_results} which refer to robustness checks and additional estimations which complement the main results presented in Section \ref{subsec:benchmark_results}. 

\begin{landscape} 
\begin{figure}[ht]
    \centering
    \caption{Impulse Response to One-Standard-Deviation Shock \\ \footnotesize Separate Samples for Geographically Close \& Further Away Countries }
    \label{fig:Benchmark_Regions}
     \centering
     \begin{subfigure}[b]{0.495\textwidth}
         \centering
         \includegraphics[width=\textwidth,height=9.5cm]{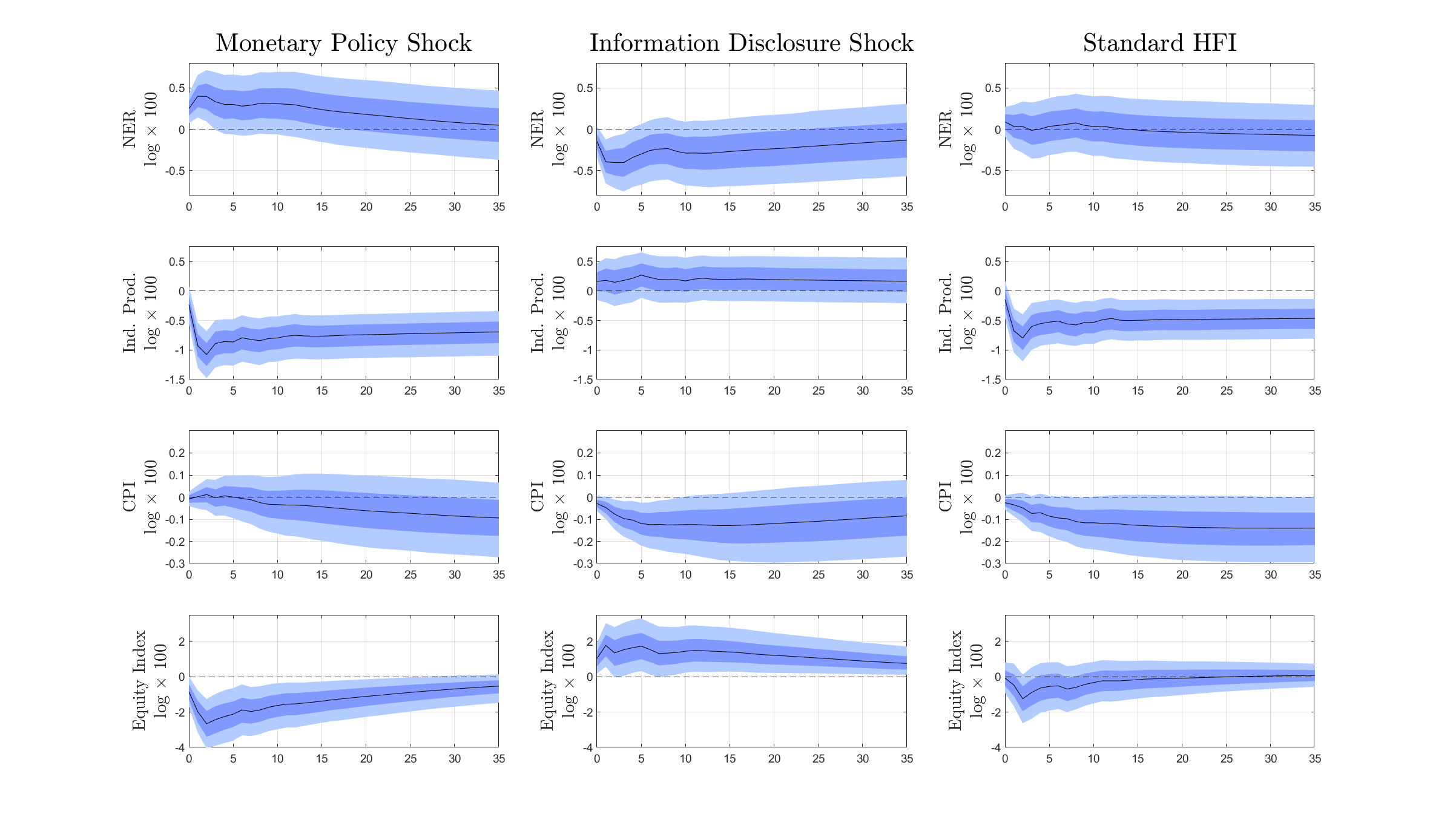}
         \caption{Closer Economies}
         \label{fig:Benchmark_CLOSER}
     \end{subfigure}
     \hfill
     \begin{subfigure}[b]{0.495\textwidth}
         \centering
         \includegraphics[width=\textwidth,height=9.5cm]{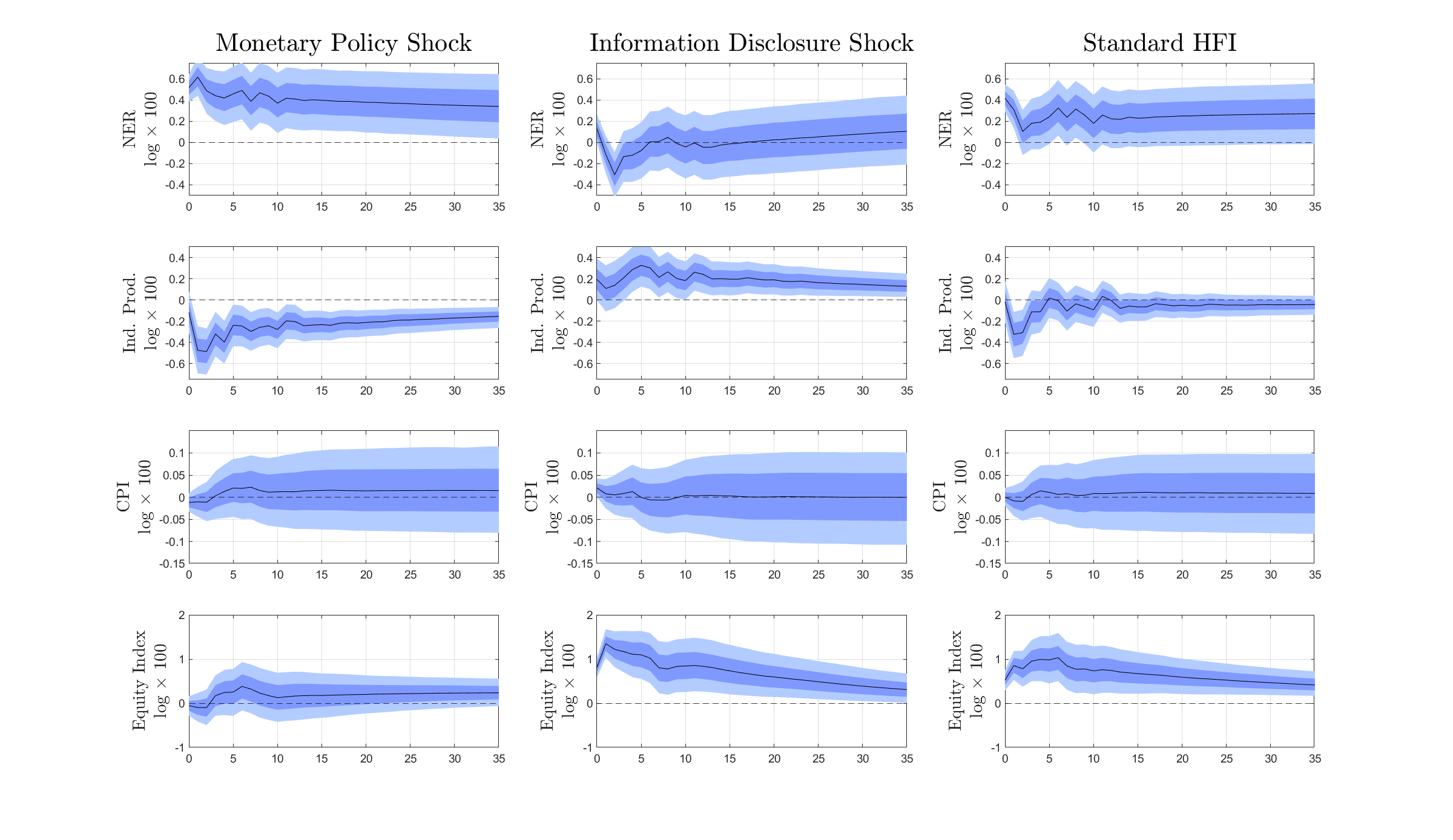}
         \caption{Further Away Economies}
         \label{fig:Benchmark_FAR}
     \end{subfigure} 
     \floatfoot{\scriptsize \textbf{Note:} The black solid line represents the median impulse response function. The dark shaded area represents the 16 and 84 percentiles. The light shaded are represents the 5 and 95 percentiles. Figure \ref{fig:Benchmark_CLOSER} on the left presents the results for the panel of ``Close Economies'' comprised of: Bulgaria, Hungary, Iceland, Poland, Russia, Sweden, Turkey. Figure \ref{fig:Benchmark_FAR} presents the results for the panel of ``Further Away Economies'' comprised of: Australia, Brazil, Canada, Chile, Colombia, India, Indonesia, Japan, Korea, Malaysia, Mexico, Peru, Philippines, Singapore, South Africa, Uruguay. Each figure is comprised of 12 sub-figures ordered in four rows and three columns. Every row represents a different variable: (i) nominal exchange rate, (ii) industrial production index, (iii) consumer price index, (iv)  equity index. The first column presents the results for the MP or ``Pure Monetary Policy'' shock, the middle column presents the results for the ID or ``Information Disclosure'' shock, and the last column presents the results for the interest rate composite high frequency surprise or ``Standard HFI''. In the text, when referring to Panel $(i,j)$, $i$ refers to the row and $j$ to the column of the figure.}
\end{figure}
\end{landscape}

\begin{landscape} 
\begin{figure}[ht]
    \centering
    \caption{Impulse Response to One-Standard-Deviation Shock \\ \footnotesize Emerging vs Advanced Economies }
    \label{fig:Benchmark_EMsVsAdv}
     \centering
     \begin{subfigure}[b]{0.495\textwidth}
         \centering
         \includegraphics[width=\textwidth,height=9.5cm]{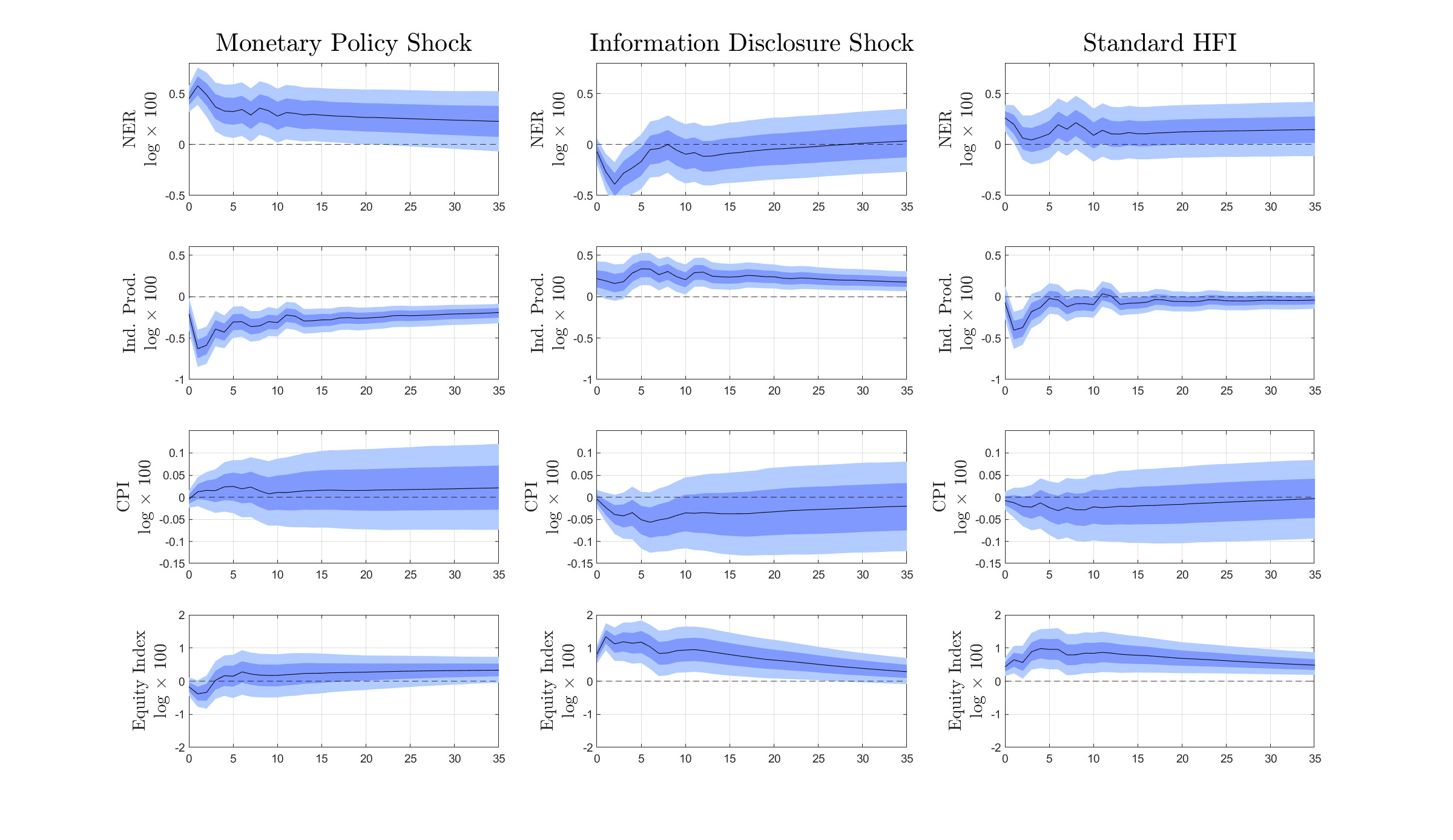}
         \caption{Emerging Economies}
         \label{fig:Benchmark_EM}
     \end{subfigure}
     \hfill
     \begin{subfigure}[b]{0.495\textwidth}
         \centering
         \includegraphics[width=\textwidth,height=9.5cm]{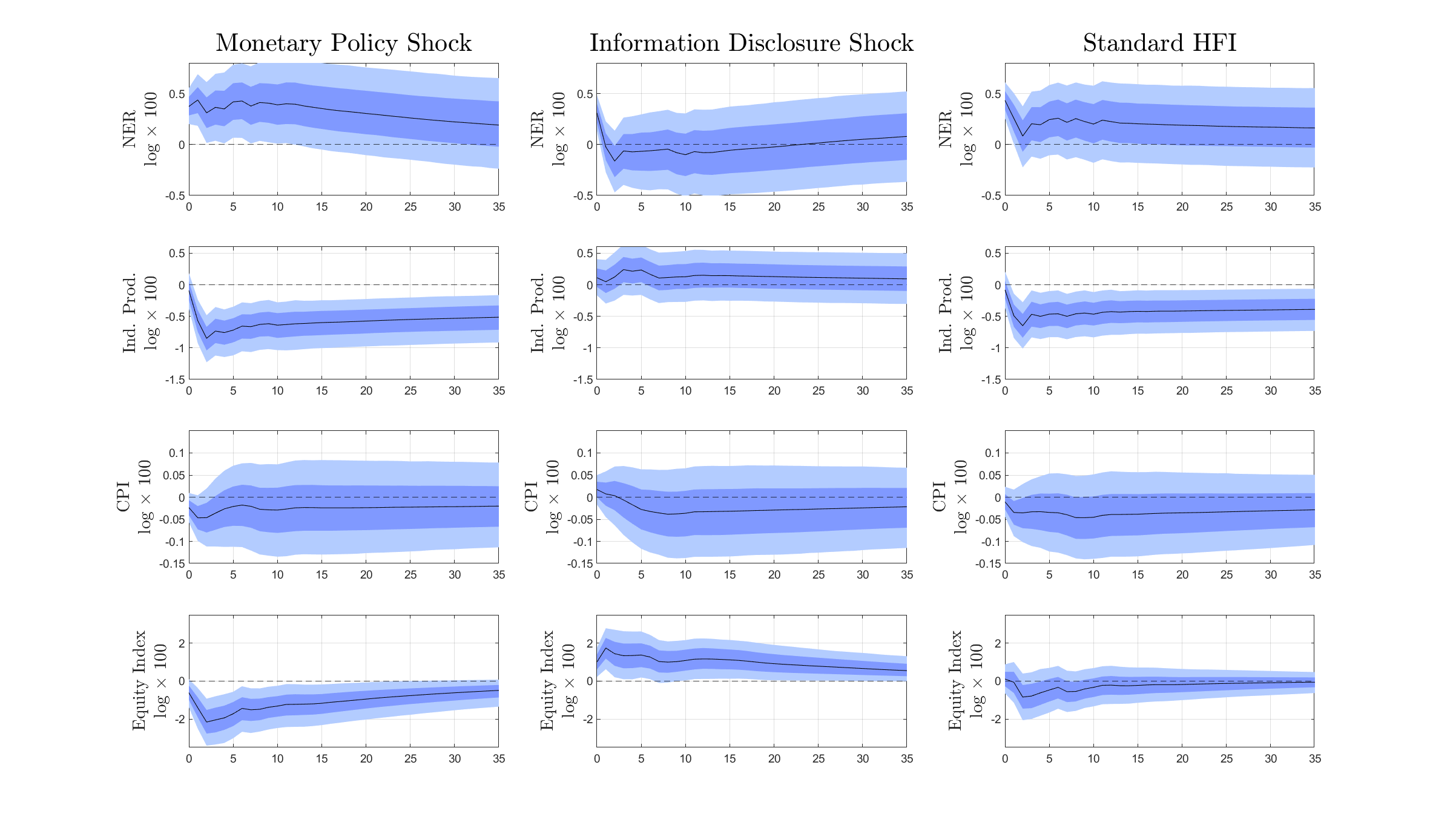}
         \caption{Advanced Economies}
         \label{fig:Benchmark_Adv}
     \end{subfigure} 
     \floatfoot{\scriptsize \textbf{Note:} The black solid line represents the median impulse response function. The dark shaded area represents the 16 and 84 percentiles. The light shaded are represents the 5 and 95 percentiles. Figure \ref{fig:Benchmark_EM} on the left presents the results for the panel of ``Emerging Economies'' comprised of: Brazil, Bulgaria, Chile, Colombia, Hungary, India, Indonesia, Malaysia, Mexico, Peru, Philippines, Poland, Russia, South Africa, Turkey, Uruguay. Figure \ref{fig:Benchmark_Adv} presents the results for the panel of ``Advanced Economies'' comprised of: Australia, Canada, Iceland, Japan, Singapore, South Korea, Sweden. Each figure is comprised of 12 sub-figures ordered in four rows and three columns. Every row represents a different variable: (i) nominal exchange rate, (ii) industrial production index, (iii) consumer price index, (iv)  equity index. The first column presents the results for the MP or ``Pure Monetary Policy'' shock, the middle column presents the results for the ID or ``Information Disclosure'' shock, and the last column presents the results for the interest rate composite high frequency surprise or ``Standard HFI''. In the text, when referring to Panel $(i,j)$, $i$ refers to the row and $j$ to the column of the figure.}
\end{figure}
\end{landscape}


\newpage
\begin{figure}[ht]
         \centering
         \caption{Local Projection Regression - Parameter $\beta^{Shock}_h$ \\ \footnotesize Full Sample}
         \includegraphics[scale=0.425]{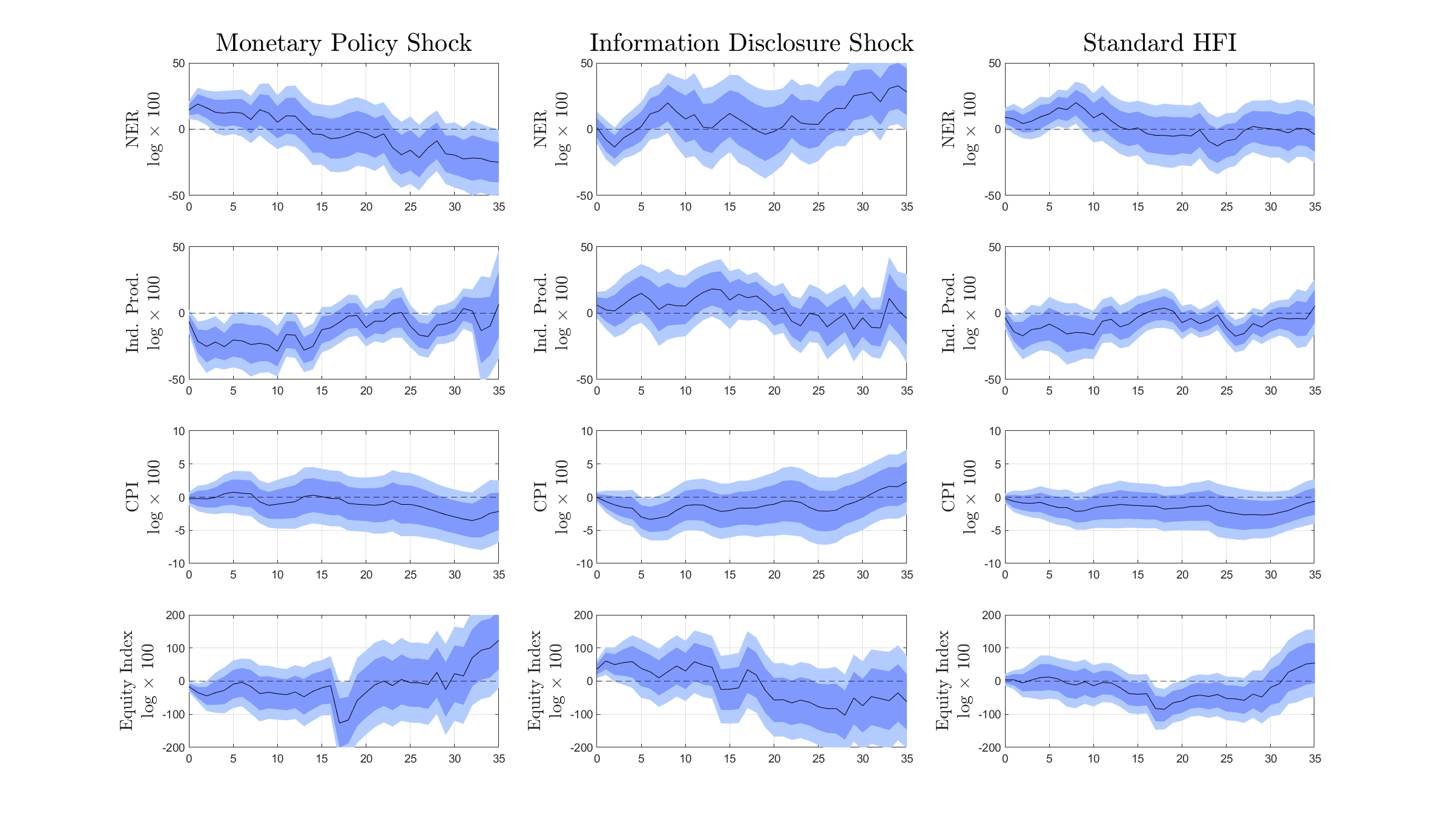}
         \label{fig:Benchmark_LP}
         \floatfoot{\scriptsize \textbf{Note:} The black solid line represents the point estimate of parameter $\beta^{Shock}_h$ from estimating Equation \ref{eq:LP_fixed_effects_date}. The dark shaded area represents the 68 confidence intervals. The light shaded area represents 90 confidence intervals. The figure is comprised of 12 sub-figures ordered in four rows and three columns. Every row represents a different variable: (i) nominal exchange rate, (ii) industrial production index, (iii) consumer price index, (iv) equity index. The first column presents the results for the MP or ``Pure Monetary Policy'' shock, the middle column presents the results for the ID or ``Information Disclosure'' shock, and the last column presents the results for the interest rate composite high frequency surprise or ``Standard HFI''. In the text, when referring to Panel $(i,j)$, $i$ refers to the row and $j$ to the column of the figure. Standard errors are clustered at the ``time'' or ``date'' level.}
\end{figure}

\begin{figure}[ht]
         \centering
         \caption{Local Projection Regression - Parameter $\beta^{Shock}_h$ \\ \footnotesize Full Sample - No Country Fixed Effects}
         \includegraphics[scale=0.425]{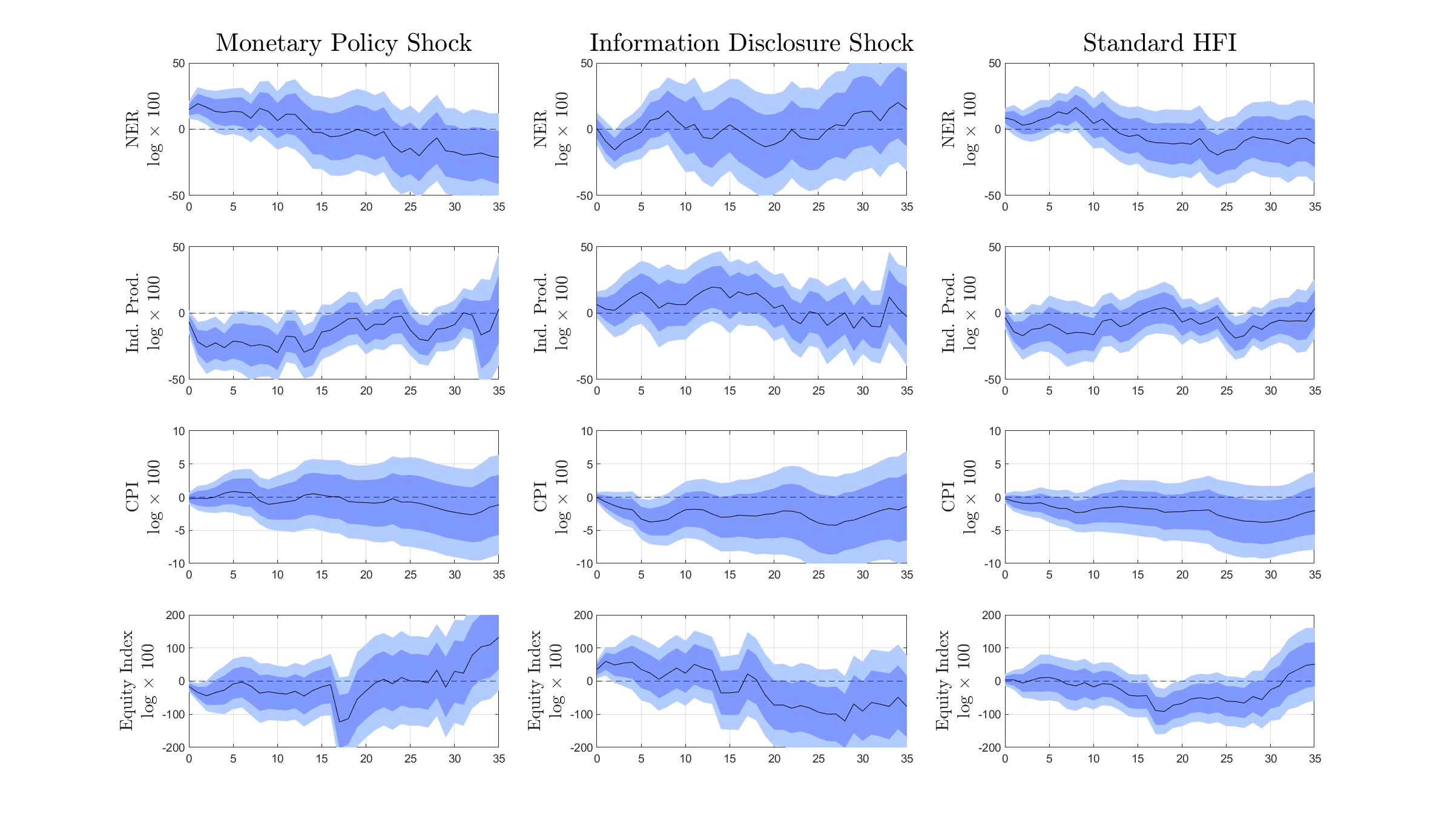}
         \label{fig:Benchmark_LP_NOFE}
         \floatfoot{\scriptsize \textbf{Note:} The black solid line represents the point estimate of parameter $\beta^{Shock}_h$ from estimating Equation \ref{eq:LP_fixed_effects_date} without the country fixed-effects $\Gamma_i$. The dark shaded area represents the 68 confidence intervals. The light shaded area represents 90 confidence intervals. The figure is comprised of 12 sub-figures ordered in four rows and three columns. Every row represents a different variable: (i) nominal exchange rate, (ii) industrial production index, (iii) consumer price index, (iv) equity index. The first column presents the results for the MP or ``Pure Monetary Policy'' shock, the middle column presents the results for the ID or ``Information Disclosure'' shock, and the last column presents the results for the interest rate composite high frequency surprise or ``Standard HFI''. In the text, when referring to Panel $(i,j)$, $i$ refers to the row and $j$ to the column of the figure. Standard errors are clustered at the ``time'' or ``date'' level.}
\end{figure}

\begin{figure}[ht]
         \centering
         \caption{Local Projection Regression - Parameter $\beta^{Shock}_h$ \\ \footnotesize Full Sample - No FE - No Time Trend}
         \includegraphics[scale=0.425]{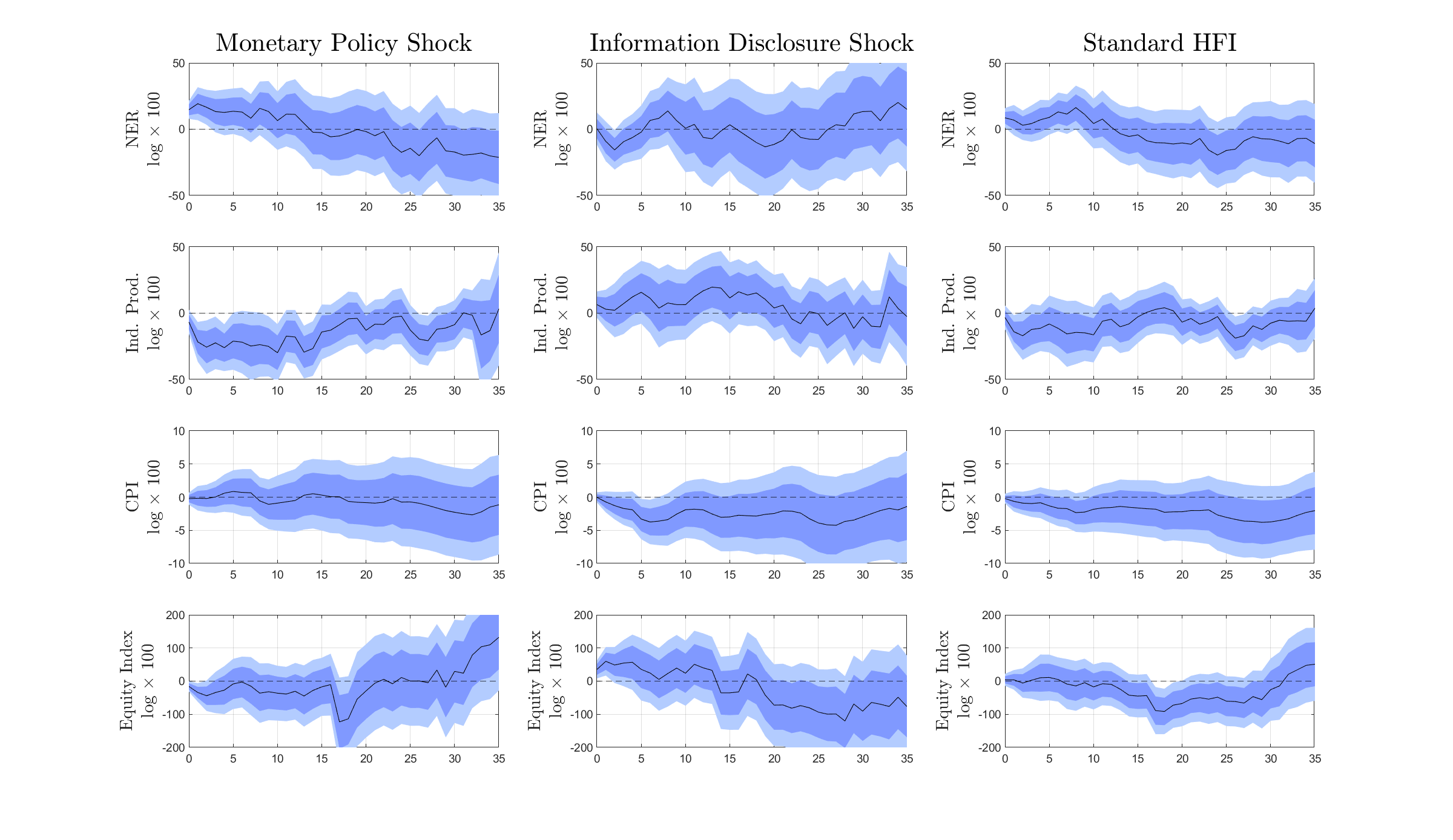}
         \label{fig:Benchmark_LP_NOFE_NoTrend}
         \floatfoot{\scriptsize \textbf{Note:} The black solid line represents the point estimate of parameter $\beta^{Shock}_h$ from estimating Equation \ref{eq:LP_fixed_effects_date} without the country fixed-effects $\Gamma_i$ and the time trend $\Gamma_{i,t}$. The dark shaded area represents the 68 confidence intervals. The light shaded area represents 90 confidence intervals. The figure is comprised of 12 sub-figures ordered in four rows and three columns. Every row represents a different variable: (i) nominal exchange rate, (ii) industrial production index, (iii) consumer price index, (iv) equity index. The first column presents the results for the MP or ``Pure Monetary Policy'' shock, the middle column presents the results for the ID or ``Information Disclosure'' shock, and the last column presents the results for the interest rate composite high frequency surprise or ``Standard HFI''. In the text, when referring to Panel $(i,j)$, $i$ refers to the row and $j$ to the column of the figure. Standard errors are clustered at the ``time'' or ``date'' level.}
\end{figure}

\begin{figure}[ht]
         \centering
         \caption{Local Projection Regression - Parameter $\beta^{Shock}_h$ \\ \footnotesize ``Close'' Sub-Sample}
         \includegraphics[scale=0.425]{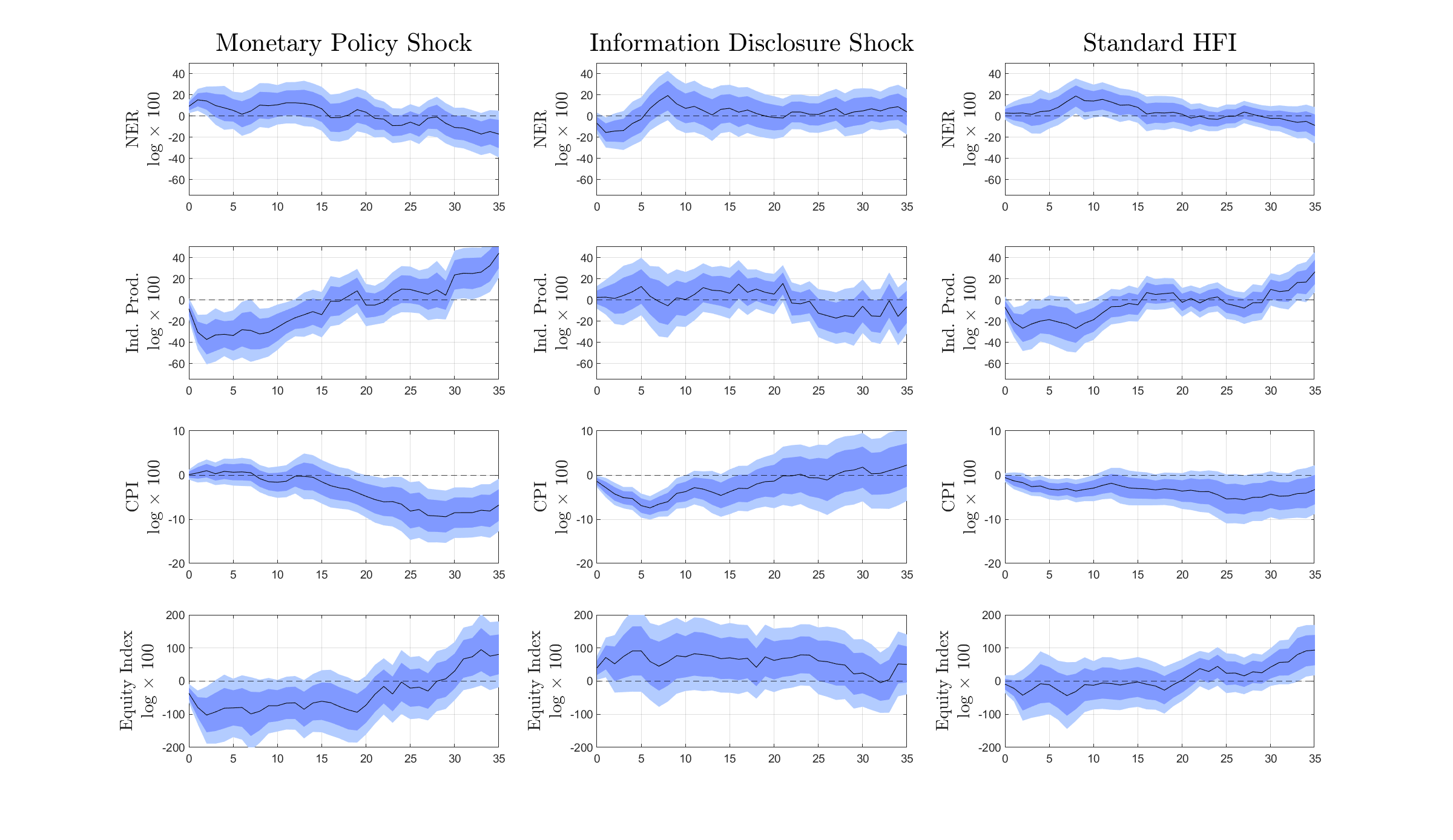}
         \label{fig:Benchmark_LP_Closer}
         \floatfoot{\scriptsize \textbf{Note:} The black solid line represents the point estimate of parameter $\beta^{Shock}_h$ from estimating Equation \ref{eq:LP_fixed_effects_date}. The dark shaded area represents the 68 confidence intervals. The light shaded area represents 90 confidence intervals. The figure is comprised of 12 sub-figures ordered in four rows and three columns. Every row represents a different variable: (i) nominal exchange rate, (ii) industrial production index, (iii) consumer price index, (iv) equity index. The first column presents the results for the MP or ``Pure Monetary Policy'' shock, the middle column presents the results for the ID or ``Information Disclosure'' shock, and the last column presents the results for the interest rate composite high frequency surprise or ``Standard HFI''. In the text, when referring to Panel $(i,j)$, $i$ refers to the row and $j$ to the column of the figure. Standard errors are clustered at the ``time'' or ``date'' level. This figure is computed for the sub-sample of ``Close Economies'' comprised of: Bulgaria, Hungary, Iceland, Poland, Russia, Sweden, Turkey.}
\end{figure}

\begin{figure}[ht]
         \centering
         \caption{Local Projection Regression - Parameter $\beta^{Shock}_h$ \\ \footnotesize ``Further Away'' Sub-Sample}
         \includegraphics[scale=0.425]{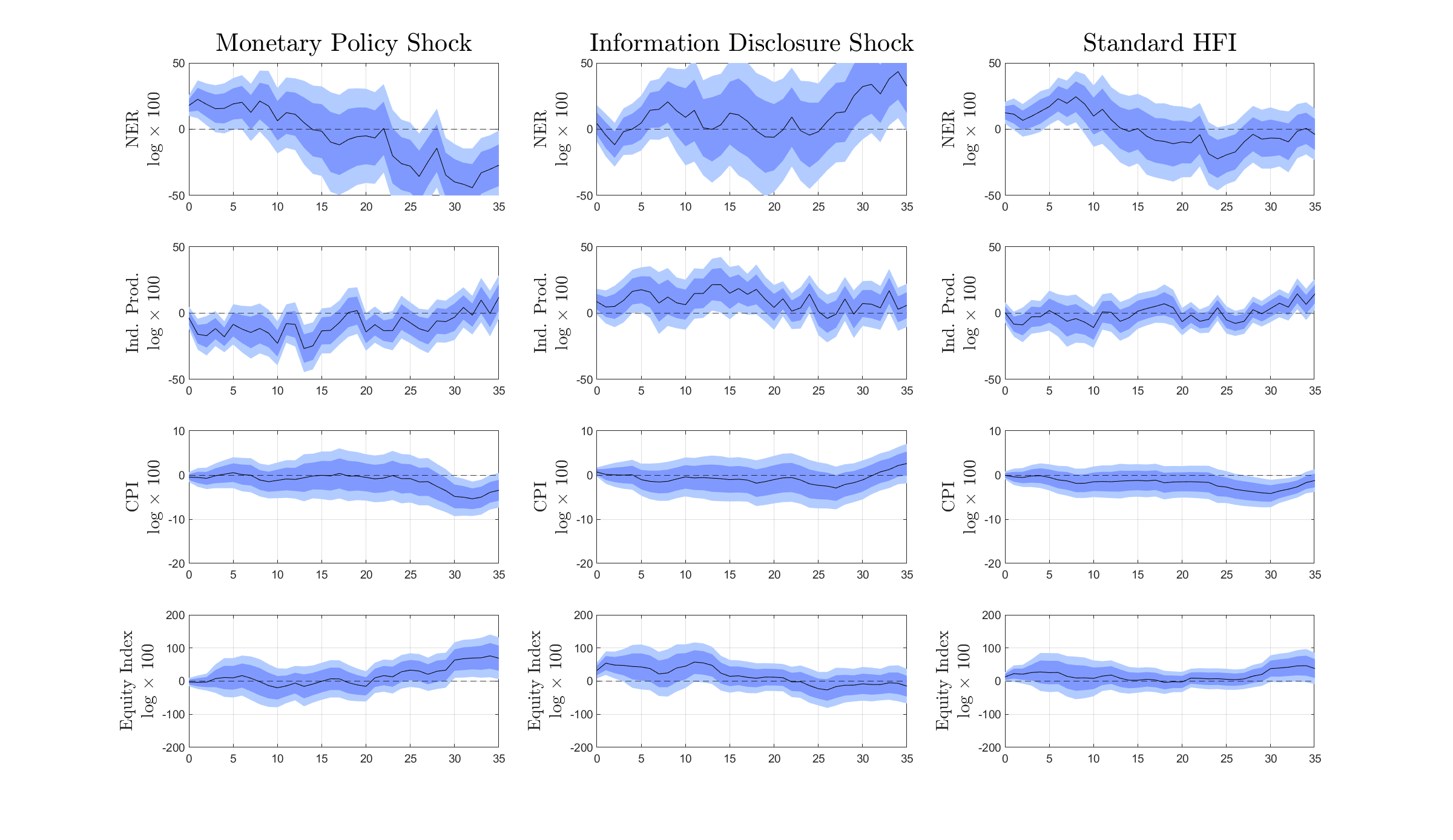}
         \label{fig:Benchmark_LP_Far}
         \floatfoot{\scriptsize \textbf{Note:} The black solid line represents the point estimate of parameter $\beta^{Shock}_h$ from estimating Equation \ref{eq:LP_fixed_effects_date}. The dark shaded area represents the 68 confidence intervals. The light shaded area represents 90 confidence intervals. The figure is comprised of 12 sub-figures ordered in four rows and three columns. Every row represents a different variable: (i) nominal exchange rate, (ii) industrial production index, (iii) consumer price index, (iv) equity index. The first column presents the results for the MP or ``Pure Monetary Policy'' shock, the middle column presents the results for the ID or ``Information Disclosure'' shock, and the last column presents the results for the interest rate composite high frequency surprise or ``Standard HFI''. In the text, when referring to Panel $(i,j)$, $i$ refers to the row and $j$ to the column of the figure. Standard errors are clustered at the ``time'' or ``date'' level. This figure is computed for the sub-sample of ``Further Away Economies'' comprised of: Australia, Brazil, Canada, Chile, Colombia, India, Indonesia, Japan, Korea, Malaysia, Mexico, Peru, Philippines, Singapore, South Africa, Uruguay.}
\end{figure}

\begin{figure}[ht]
         \centering
         \caption{Local Projection Regression - Parameter $\beta^{Shock}_h$ \\ \footnotesize Emerging Economies Sub-Sample}
         \includegraphics[scale=0.425]{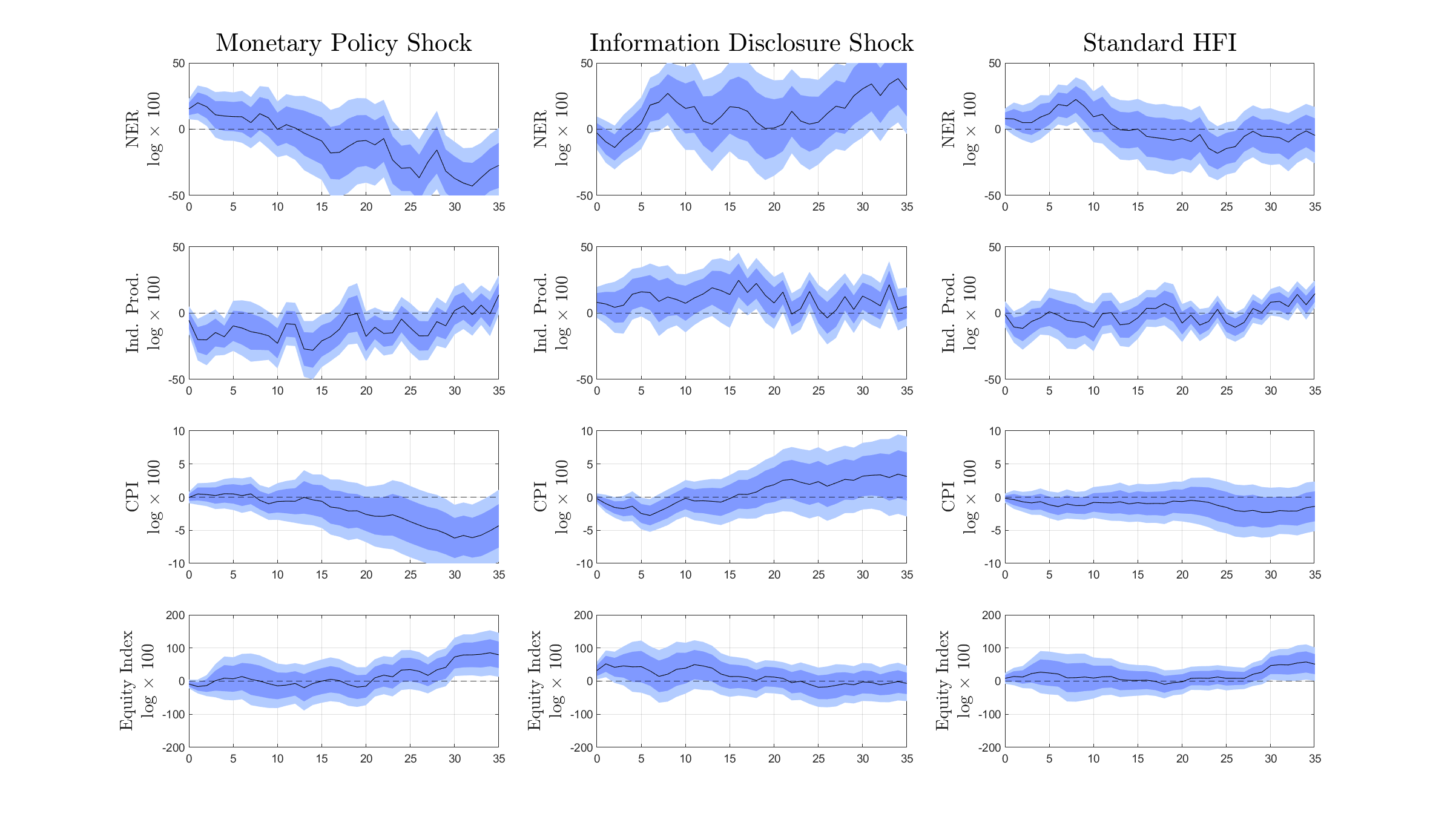}
         \label{fig:Benchmark_LP_EM}
         \floatfoot{\scriptsize \textbf{Note:} The black solid line represents the point estimate of parameter $\beta^{Shock}_h$ from estimating Equation \ref{eq:LP_fixed_effects_date}. The dark shaded area represents the 68 confidence intervals. The light shaded area represents 90 confidence intervals. The figure is comprised of 12 sub-figures ordered in four rows and three columns. Every row represents a different variable: (i) nominal exchange rate, (ii) industrial production index, (iii) consumer price index, (iv) equity index. The first column presents the results for the MP or ``Pure Monetary Policy'' shock, the middle column presents the results for the ID or ``Information Disclosure'' shock, and the last column presents the results for the interest rate composite high frequency surprise or ``Standard HFI''. In the text, when referring to Panel $(i,j)$, $i$ refers to the row and $j$ to the column of the figure. Standard errors are clustered at the ``time'' or ``date'' level. This figure is computed for the sub-sample of ``Emerging Market Economies'' comprised of: Brazil, Bulgaria, Chile, Colombia, Hungary, India, Indonesia, Malaysia, Mexico, Peru, Philippines, Poland, Russia, South Africa, Turkey, Uruguay.}
\end{figure}

\begin{figure}[ht]
         \centering
         \caption{Local Projection Regression - Parameter $\beta^{Shock}_h$ \\ \footnotesize Advanced Economies Sub-Sample}
         \includegraphics[scale=0.425]{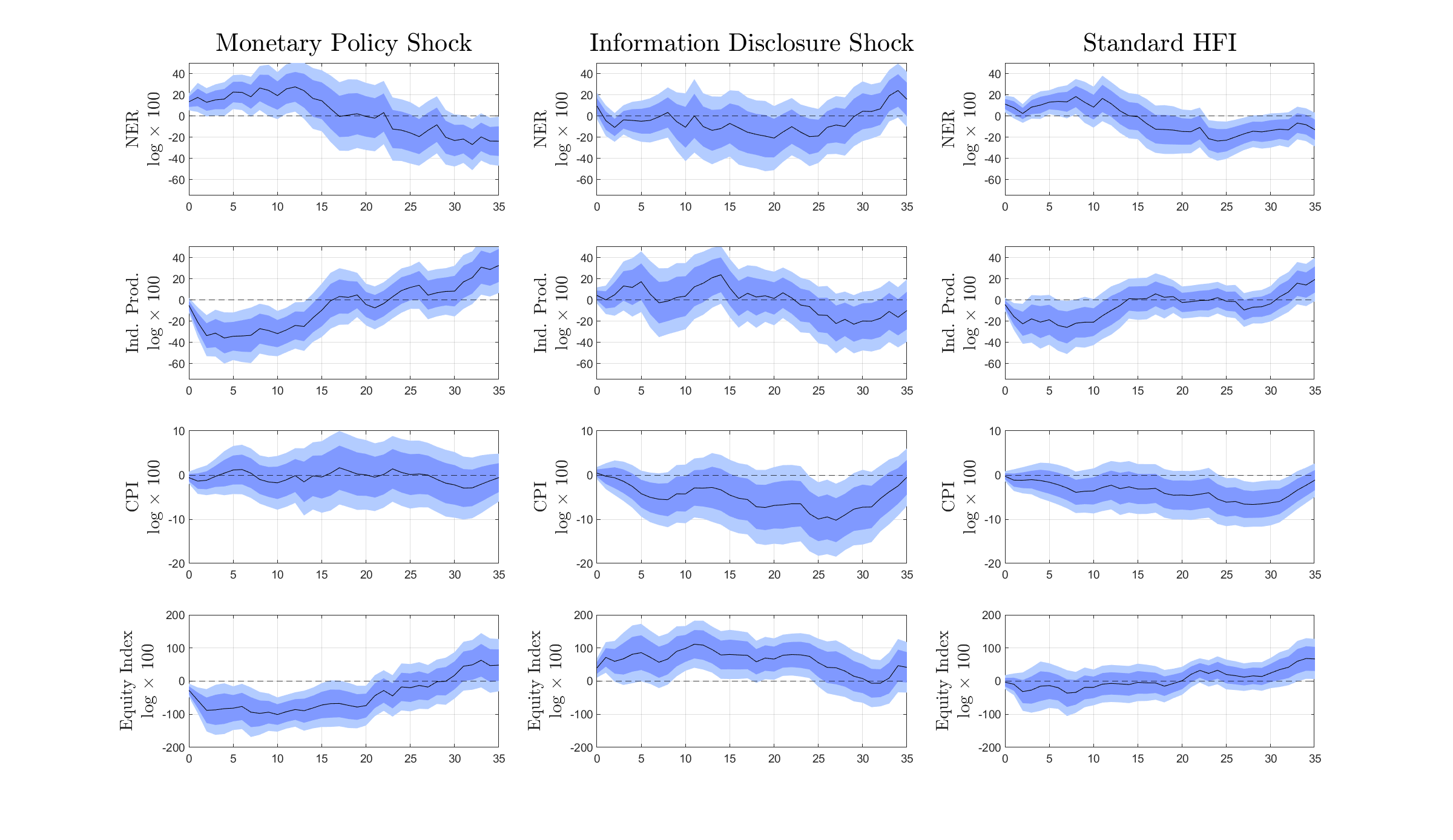}
         \label{fig:Benchmark_LP_Adv}
         \floatfoot{\scriptsize \textbf{Note:} The black solid line represents the point estimate of parameter $\beta^{Shock}_h$ from estimating Equation \ref{eq:LP_fixed_effects_date}. The dark shaded area represents the 68 confidence intervals. The light shaded area represents 90 confidence intervals. The figure is comprised of 12 sub-figures ordered in four rows and three columns. Every row represents a different variable: (i) nominal exchange rate, (ii) industrial production index, (iii) consumer price index, (iv) equity index. The first column presents the results for the MP or ``Pure Monetary Policy'' shock, the middle column presents the results for the ID or ``Information Disclosure'' shock, and the last column presents the results for the interest rate composite high frequency surprise or ``Standard HFI''. In the text, when referring to Panel $(i,j)$, $i$ refers to the row and $j$ to the column of the figure. Standard errors are clustered at the ``time'' or ``date'' level. This figure is computed for the sub-sample of ``Emerging Market Economies'' comprised of: Australia, Canada, Iceland, Japan, Singapore, South Korea, Sweden.}
\end{figure}

\newpage
\begin{figure}[ht]
         \centering
         \caption{Impulse Response to One-Standard-Deviation Shock \\ \footnotesize Poor Man Rotational Angle}
         \includegraphics[scale=0.425]{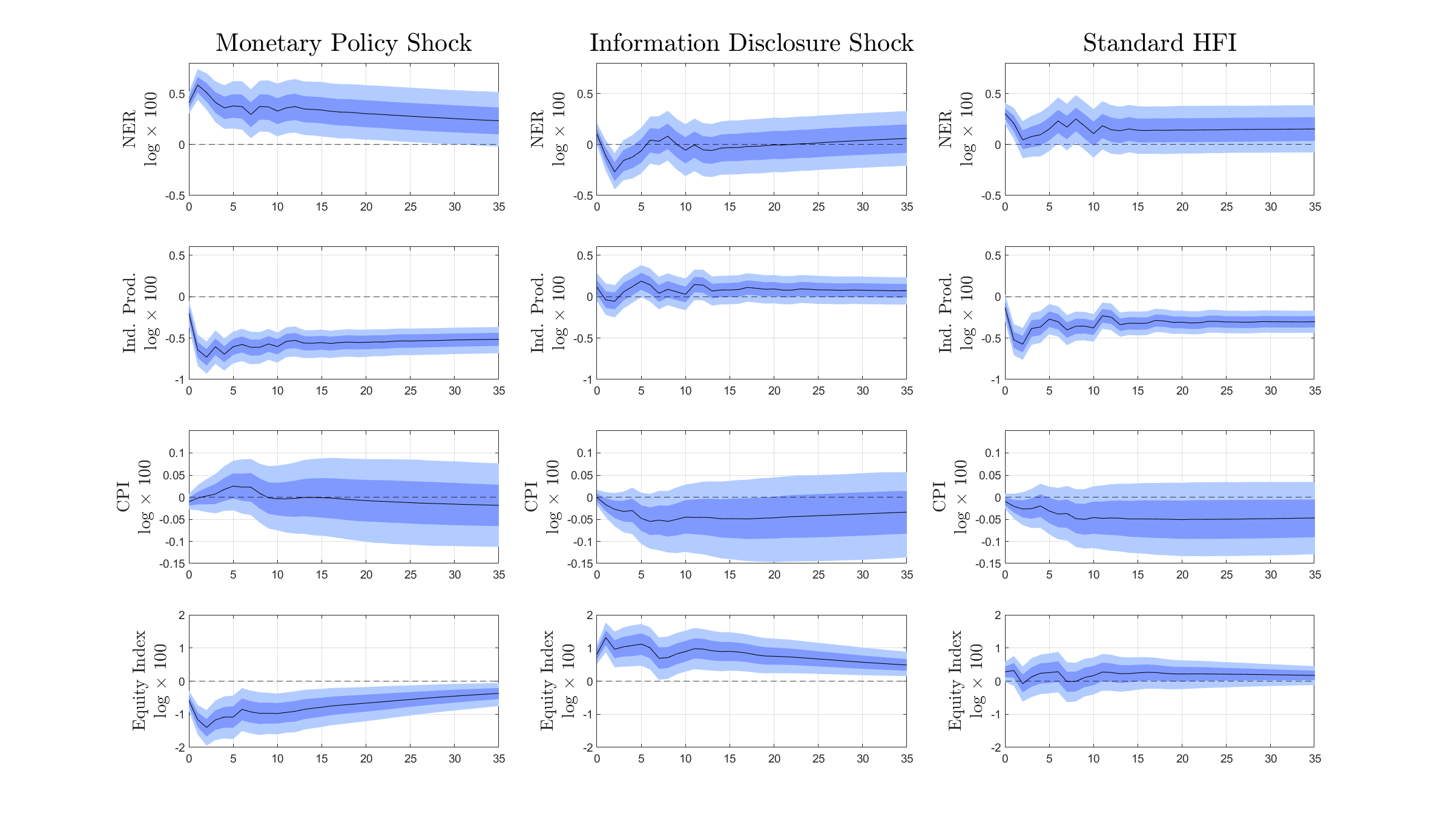}
         \label{fig:Benchmark_j}
         \floatfoot{\scriptsize \textbf{Note:} The black solid line represents the median impulse response function. The dark shaded area represents the 68 confidence intervals. The light shaded area represents 95 confidence intervals. The figure is comprised of 12 sub-figures ordered in four rows and three columns. Every row represents a different variable: (i) nominal exchange rate, (ii) industrial production index, (iii) consumer price index, (iv) equity index. The first column presents the results for the MP or ``Pure Monetary Policy'' shock, the middle column presents the results for the ID or ``Information Disclosure'' shock, and the last column presents the results for the interest rate composite high frequency surprise or ``Standard HFI''. In the text, when referring to Panel $(i,j)$, $i$ refers to the row and $j$ to the column of the figure.}
\end{figure}

\begin{figure}[ht]
         \centering
         \caption{Impulse Response to One-Standard-Deviation Shock \\ \footnotesize $10^{th}$ Percentile Rotational Angle}
         \includegraphics[scale=0.425]{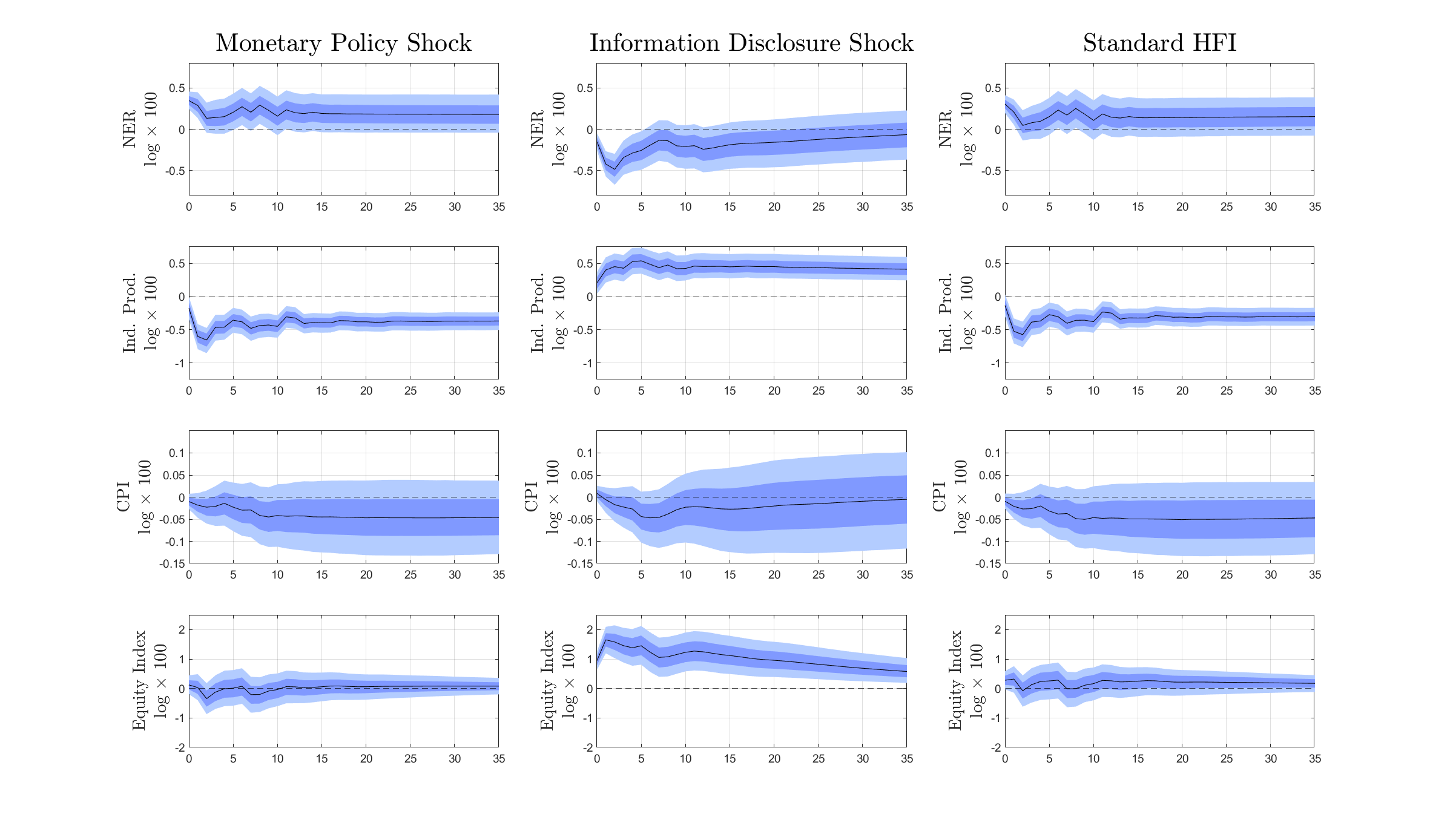}
         \label{fig:Benchmark_10}
         \floatfoot{\scriptsize \textbf{Note:} The black solid line represents the median impulse response function. The dark shaded area represents the 68 confidence intervals. The light shaded area represents 95 confidence intervals. The figure is comprised of 12 sub-figures ordered in four rows and three columns. Every row represents a different variable: (i) nominal exchange rate, (ii) industrial production index, (iii) consumer price index, (iv) equity index. The first column presents the results for the MP or ``Pure Monetary Policy'' shock, the middle column presents the results for the ID or ``Information Disclosure'' shock, and the last column presents the results for the interest rate composite high frequency surprise or ``Standard HFI''. In the text, when referring to Panel $(i,j)$, $i$ refers to the row and $j$ to the column of the figure.}
\end{figure}

\begin{figure}[ht]
         \centering
         \caption{Impulse Response to One-Standard-Deviation Shock \\ \footnotesize $10^{th}$ Percentile Rotational Angle}
         \includegraphics[scale=0.425]{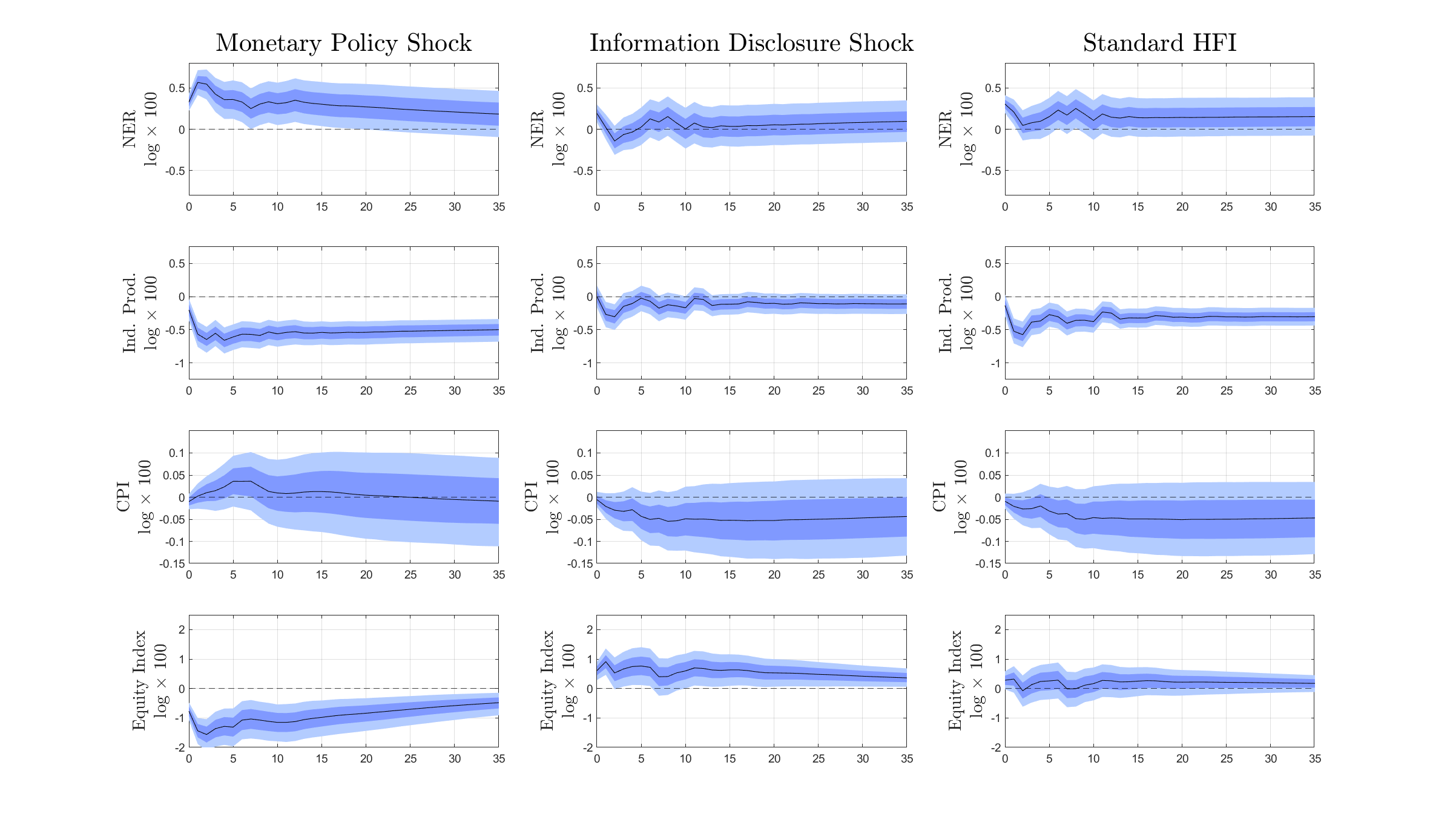}
         \label{fig:Benchmark_90}
         \floatfoot{\scriptsize \textbf{Note:} The black solid line represents the median impulse response function. The dark shaded area represents the 68 confidence intervals. The light shaded area represents 95 confidence intervals. The figure is comprised of 12 sub-figures ordered in four rows and three columns. Every row represents a different variable: (i) nominal exchange rate, (ii) industrial production index, (iii) consumer price index, (iv) equity index. The first column presents the results for the MP or ``Pure Monetary Policy'' shock, the middle column presents the results for the ID or ``Information Disclosure'' shock, and the last column presents the results for the interest rate composite high frequency surprise or ``Standard HFI''. In the text, when referring to Panel $(i,j)$, $i$ refers to the row and $j$ to the column of the figure.}
\end{figure}

\begin{figure}[ht]
         \centering
         \caption{Impulse Response to One-Standard-Deviation Shock \\ \footnotesize Alternative Identification Strategy}
         \includegraphics[scale=0.425]{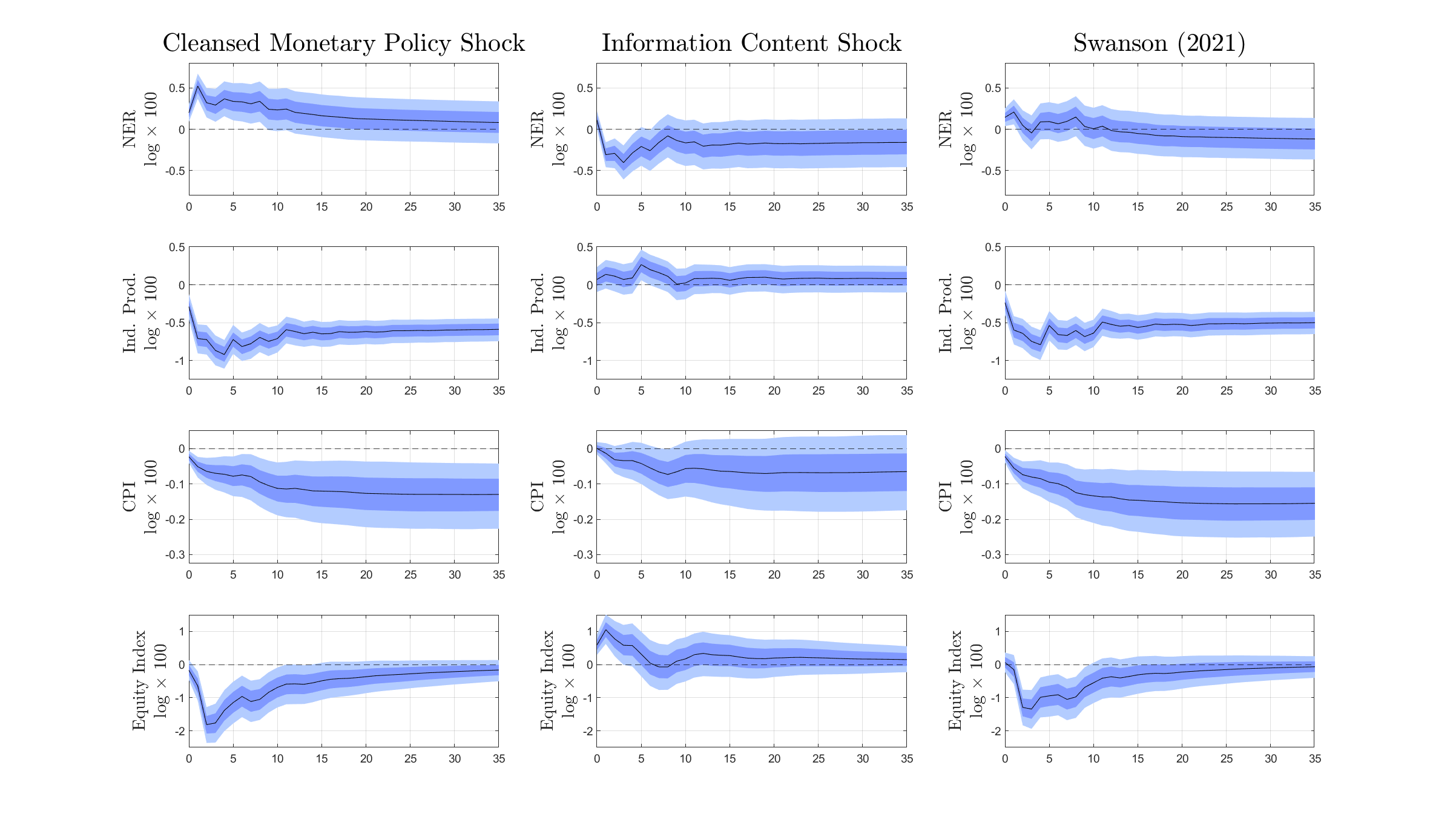}
         \label{fig:Alternative_Identification}
         \floatfoot{\scriptsize \textbf{Note:} The black solid line represents the median impulse response function. The dark shaded area represents the 68 confidence intervals. The light shaded area represents 95 confidence intervals. The figure is comprised of 12 sub-figures ordered in four rows and three columns. Every row represents a different variable: (i) nominal exchange rate, (ii) industrial production index, (iii) consumer price index, (iv) equity index. The first column presents the results for the ``Cleansed Monetary Policy'' shock, the middle column presents the results for the ID or ``Information Content'' shock, and the last column presents the results for the interest rate composite high frequency surprise constructed by \cite{swanson2021measuring}. In the text, when referring to Panel $(i,j)$, $i$ refers to the row and $j$ to the column of the figure.}
\end{figure}

\begin{figure}[ht]
         \centering
         \caption{Local Projection Regression - Parameter $\beta^{Shock}_h$ \\ \footnotesize Alt. Identification Strategy}
         \includegraphics[scale=0.425]{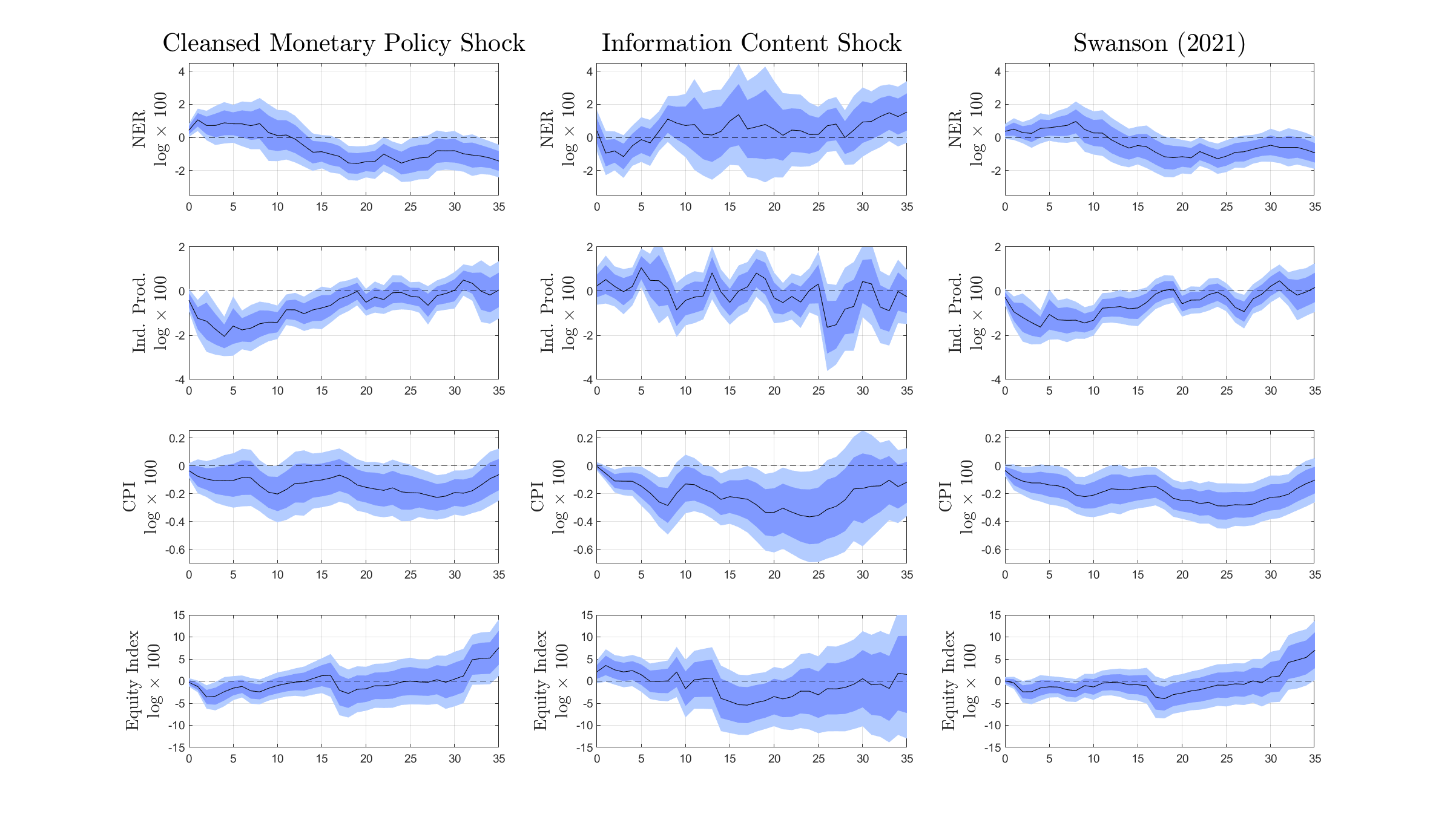}
         \label{fig:Alternative_LP}
         \floatfoot{\scriptsize \textbf{Note:} The black solid line represents the point estimate of parameter $\beta^{Shock}_h$ from estimating Equation \ref{eq:LP_fixed_effects_date}. The dark shaded area represents the 68 confidence intervals. The light shaded area represents 90 confidence intervals. The figure is comprised of 12 sub-figures ordered in four rows and three columns. Every row represents a different variable: (i) nominal exchange rate, (ii) industrial production index, (iii) consumer price index, (iv) equity index. The first column presents the results for the ``Cleansed Monetary Policy'' shock, the middle column presents the results for the ID or ``Information Content'' shock, and the last column presents the results for the interest rate composite high frequency surprise constructed by \cite{swanson2021measuring}. In the text, when referring to Panel $(i,j)$, $i$ refers to the row and $j$ to the column of the figure.}
\end{figure}

\begin{figure}[ht]
         \centering
         \caption{Local Projection Regression - Parameter $\beta^{Shock}_h$ \\ \footnotesize Rate Race}
         \includegraphics[scale=0.425]{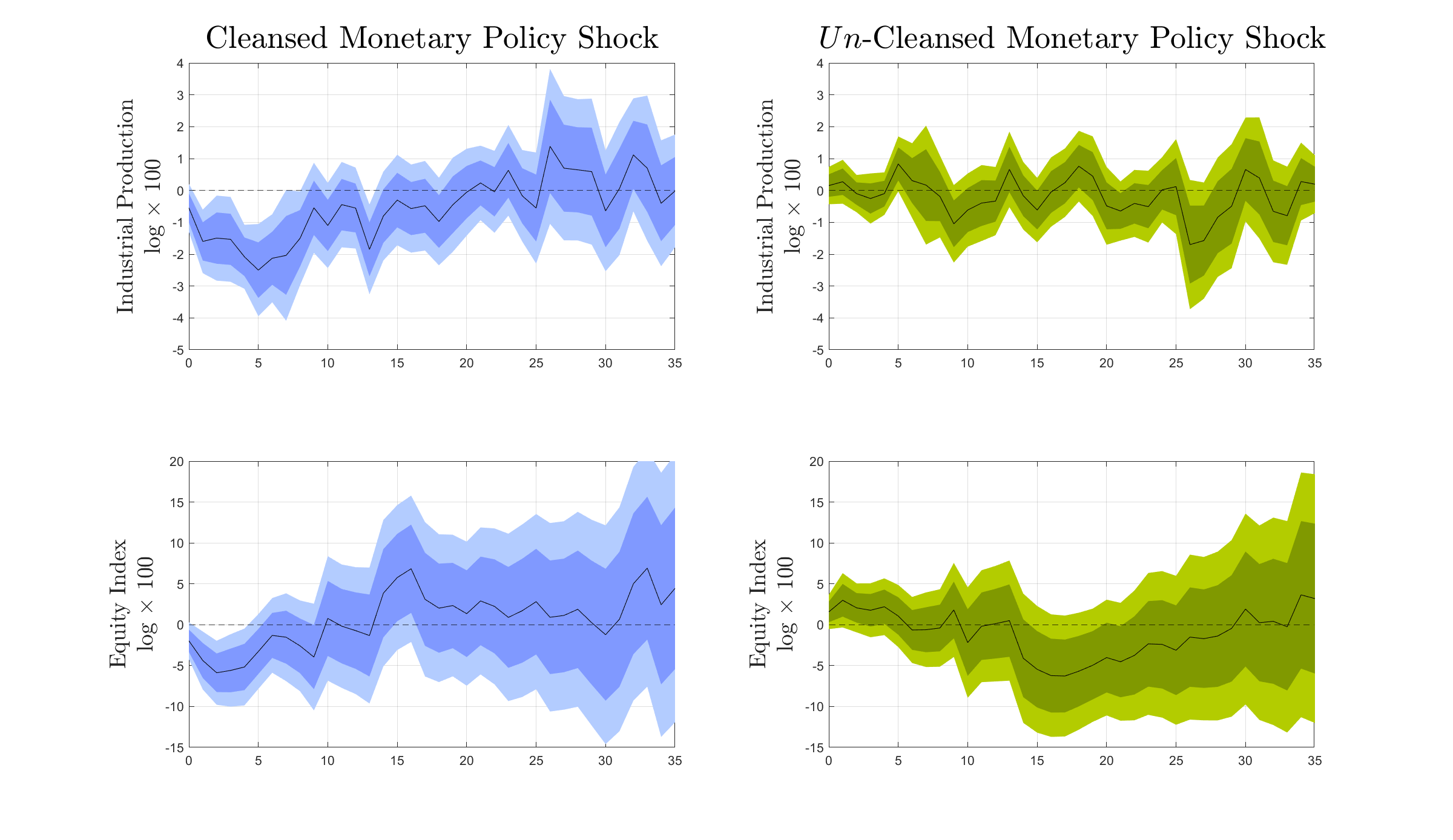}
         \label{fig:Rate_Race_LP}
         \floatfoot{\scriptsize \textbf{Note:} The black solid line represents the point estimate of parameter $\beta^{Shock}_h$ from estimating Equation \ref{eq:LP_fixed_effects_date} with two shocks simultaneously. The dark shaded area represents the 68 confidence intervals. The light shaded area represents 90 confidence intervals. The figure is comprised of 4 sub-figures ordered in two rows and two columns. Every row represents a different variable: (i) industrial production index, (ii) equity index. The first column presents the results for the ``Cleansed Monetary Policy'' shock, the second column presents the results for the ID or ``Uncleansed'' or \cite{swanson2021measuring} shock. In the text, when referring to Panel $(i,j)$, $i$ refers to the row and $j$ to the column of the figure.}
\end{figure}

\end{document}